\documentclass[useAMS,usenatbib,usegraphics]{mn2e}

 \usepackage{times}
 \usepackage{amssymb}
 \usepackage{graphicx}
 \usepackage{natbib}

\title[Stellar populations in the third quadrant of the Galaxy]
{The complex stellar populations in the lines of sight to open clusters in the third Galactic quadrant} 

\author[Carraro, Seleznev, Baume and Turner]{Giovanni Carraro,$^{1}$\thanks{On leave of absence from Padova University.E-mail:gcarraro@eso.org} \
Anton F. Seleznev, $^{2}$ \
Gustavo Baume,$^{3}$ \
David.~G.~Turner$^{4}$ \\
$^{1}$ESO, Alonso de Cordova 3107, 19001, Santiago de Chile, Chile\\
$^{2}$Astronomical Observatory, Ural Federal University, Mira str. 19, Ekaterinburg, 620002, Russia\\
$^{3}$ Facultad de Ciencias Astron\'omicas y Geof\'isicas (UNLP), Universidad de La Plata (CONICET, UNLP), Paseo del Bosque s/n, La Plata, Argentina\\
$^{4}$ Department of Astronomy and Physics, Saint Mary's University, Halifax, NS~B3H~3C3, Canada}

\date{Accepted XXX. Received YYY; in original form ZZZ}

\pubyear{2015}

\begin{document}
\maketitle

\begin{abstract}
Multi-color photometry of the stellar populations in five fields in the third Galactic quadrant centred on the clusters NGC~2215, NGC~2354, Haffner~22, Ruprecht~11, and ESO489~SC01 is interpreted in terms of a warped and flared Galactic disk, without resort to an external entity such as the popular Monoceros or Canis Major overdensities. Except for NGC~2215, the clusters are poorly or unstudied previously. The data generate basic parameters for each cluster, including the distribution of stars along the line of sight. We use star counts and photometric analysis, without recourse to Galactic-model-based predictions or interpretations, and confirms earlier results for NGC~2215 and NGC~2354. ESO489~SC01 is not a real cluster, while Haffner~22 is  an overlooked  cluster aged $\sim$ 2.5 Gyr. Conclusions for Ruprecht~11 are preliminary, evidence for a  cluster being marginal. Fields surrounding the clusters show signatures of young and intermediate-age stellar populations. The young population background to NGC~2354 and Ruprecht~11 lies $\sim$8--9 kpc from the Sun and $\sim$1 kpc below the formal Galactic plane, tracing a portion of the  Norma-Cygnus arm, challenging Galactic models that adopt a sharp cut-off of the disk 12$-$14 kpc from the Galactic center. The old population is metal poor with an age of $\sim2--3$ Gyr, resembling star clusters like Tombaugh~2 or NGC~2158. It has a large color spread and is difficult to locate precisely. Young and old populations follow a pattern that depends critically on the vertical location of the thin and/or thick disk, and whether or not a particular line of sight intersects one, both, or none.
\end{abstract}
 

\section[]{Introduction}
There is a long tradition of using Galactic open clusters to study the structure and evolution of the Milky Way disk. Such ensembles contain stars with common properties, making it possible to establish solid estimates for their basic parameters (metallicity, age, and distance) in statistical fashion (Netopil et al. 2015). Relationships between parameters, such as the age-metallicity relation, the radial and vertical metallicity gradients, and so forth, make it possible to characterise the disk, and understand how it formed and evolved (Magrini et al. 2015).

The simple spatial distribution of star clusters permits us to derive global structural properties for the disk, such as the radial and vertical scale-lengths, the disk extent, and its spiral structure (Popova \& Loktin 2005, 2008; Vazquez et al. 2008; Baume et al. 2009; Carraro et al. 2010, 2015). Over the years such studies have generated a picture of the Milky Way disk that is becoming more and more solid, including: (1) the location and shape of the spiral arms close to the Sun (Perseus, Orion, and Carina-Sagittarius), (2) evidence that the outer disk is warped and flared, and, finally, (3) evidence that the Galactic disk does not have a clear density and luminosity cut-off, but most probably only a break, in full analogy with spiral galaxies in the local universe (Laine et al. 2014).

A novel approach to such studies was recently developed, beginning with early studies of the third Galactic quadrant ($180^{o} \leq l \leq 270^{o}$) by Carraro et al. (2005) and Moitinho et al. (2006). It consists of analysing the distribution of stars not only in well-known star clusters, but also in their surrounding background and forground fields. That is made possible by the structure and geometry of the Galactic disk in the third quadrant, which displays a significant warp. While the warp was noticed many years ago (Vogt \& Moffat 1975; Carney \& Seitzer 1991), it is only recently that the implications have been considered properly (Momany et al. 2006). Essentially, because of significant disk warping in the third quadrant, there are lines of sight that may completely cross the thin and thick disk. It is particularly important for the young thin disk, which, once intersected, produces prominent features in color-magnitude diagrams for the field , popularised as {\it blue plumes}. The nature of the features (color width, magnitude, tilting, and so forth) depends on the line of sight in a way that makes it possible in theory to reconstruct the properties of the thin disk by viewing in different Galactic directions (Carraro 2015).

Such a scenario has recently become more complex as a result of two additional evidences. First, the thick disk is also warped, and that produces additional features in color-magnitude diagrams that, in the past, were often interpreted as evidence of an extragalactic population (Canis Major and/or Monoceros). Second, the thick and thin disk are also flared (Lopez-Corredoira \& Molg\'o 2014), implying that, occasionally, one can find young populations far from the expected disk locus (Carraro et al. 2015; Kalberla et al. 2014; Feast et al. 2014). It is therefore important to explore new directions in such studies. 

We have adopted the strategy of centering our fields on cataloged open clusters (Dias et al. 2002). It is particularly efficient and successful since it permits us first of all to derive properties of previously unstudied groups, while at the same time providing an indication of the quality of our data, since in several cases the fields we select contain clusters for which some data are available for comparison. From such considerations we selected for study five fields strategically located in the third quadrant, and present the results here. The identity of the fields and their equatorial and Galactic coordinates are listed in Table~1.

Readers may note the extreme nature of the lines of sight selected, since they cover a broad region of the third Galactic quadrant, both in longitude and in latitude. One field lying well below the Galactic plane (ESO489~SC01) was selected in an attempt to find a direction where contamination by the thin and thick disks is negligible. A line of sight above the plane (Haffner~22) and opposite with respect to the formal Galactic plane (ESO489~SC01) was also chosen, since that is additionally where one expects minimal contamination by thin or thick disk populations. The other three fields (Ruprecht~11, NGC~2215, and NGC~2354) span a Galactic latitude range where one expects to find important but different contaminations by stars belonging to the young thin disk, the old thin disk, and perhaps also the thick disk. Except for the star cluster NGC~2215, the fields have never received proper attention previously. 

The five selected fields turn out to be very complex, and identification of the different stellar populations has been difficult. To tackle the challenge as efficiently as possible, we make use here of a combination of star counts and photometric analysis in order to separate different populations and establish their intrinsic properties (distance, reddening, and age) and spatial distributions. Our approach is not tied to any Galactic model, in contrast with studies linked to the Besancon (Robin et al. 2003) or Trilegal (Giradi et al. 2005) models, which assume an {\it a priori} description of the Milky Way, and which led to erroneous results  in the past. Among them we recall the claim for a sharp cut-off in the light and mass distribution along the Galactic plane (Robin et al. 1992, revised by Carraro et al. 2010), and the interpretation of the Canis Major over density as a dwarf galaxy ( Bellazzini et al. 2004, revised by Moitinho et al. 2006 and Momany et al. 2006).

We believe that the present study shows clearly that features belonging to the thin and thick disk are not evenly distributed across the third Galactic quadrant, but follow a pattern where the thin disk is warped and organised in spiral arms (Orion, Perseus, and Norma-Cygnus), in confirmation of our earlier results (Carraro et al. 2005; Moitinho et al. 2006).

The layout of the present work is as follows. Sect.~2 presents an overview of the literature and earlier results, if any, for the clusters centered in our fields. Sect.~3 illustrates how we collected and reduced the data, including information on calibration, merging with 2MASS, and astrometry. A preliminary discussion of the color-magnitude diagrams is provided in Sect.~4. In Sect.~5 we make use of star counts and photometric diagrams to provide estimates of each cluster's radius, age, distance, and reddening. We then discuss the various stellar populations in the fields in Sect~. 6. Finally, Sect.~7 summarises our findings.

   \begin{figure*}
   \centering
  \includegraphics[scale=0.35]{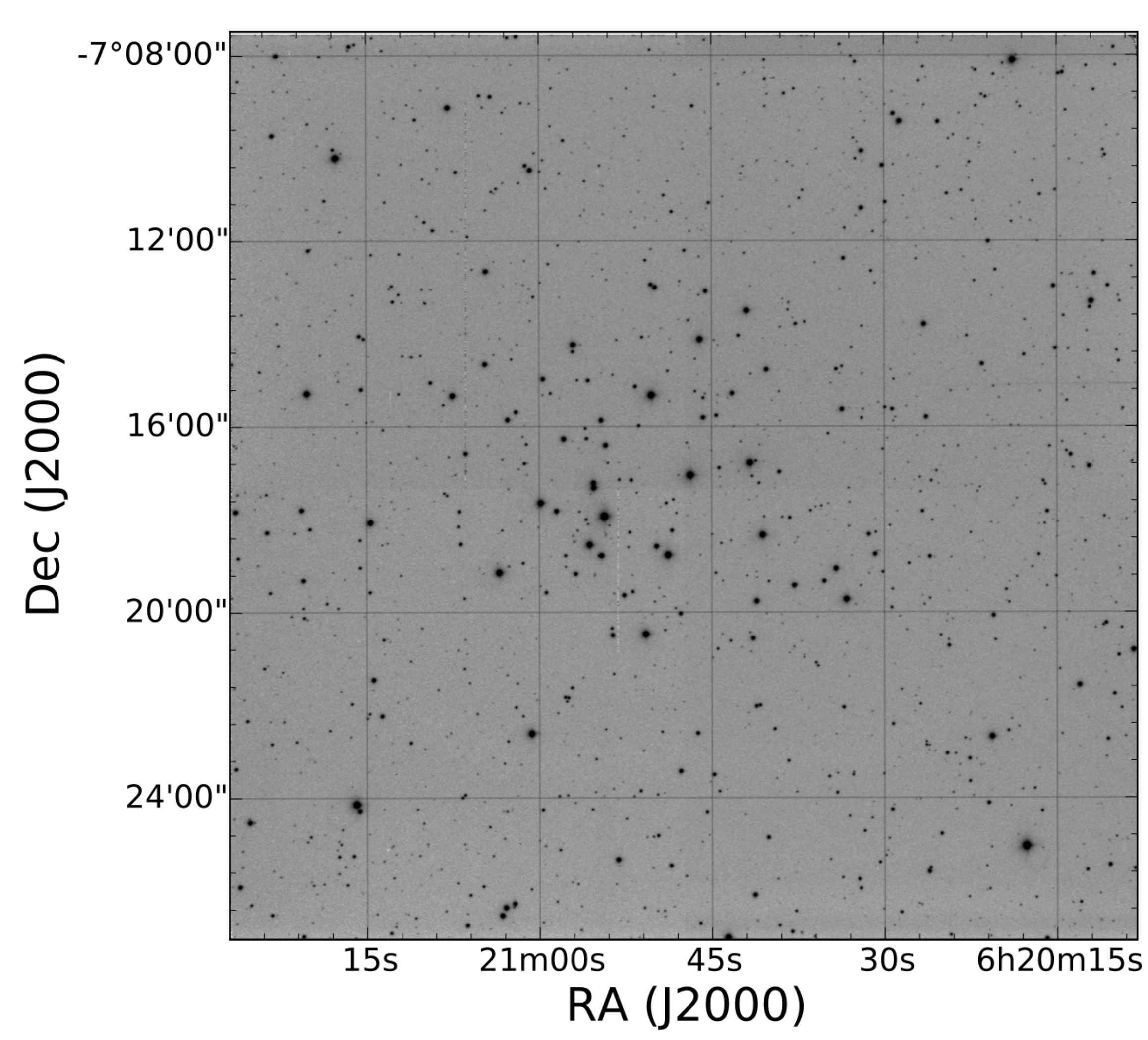}
   \includegraphics[scale=0.35]{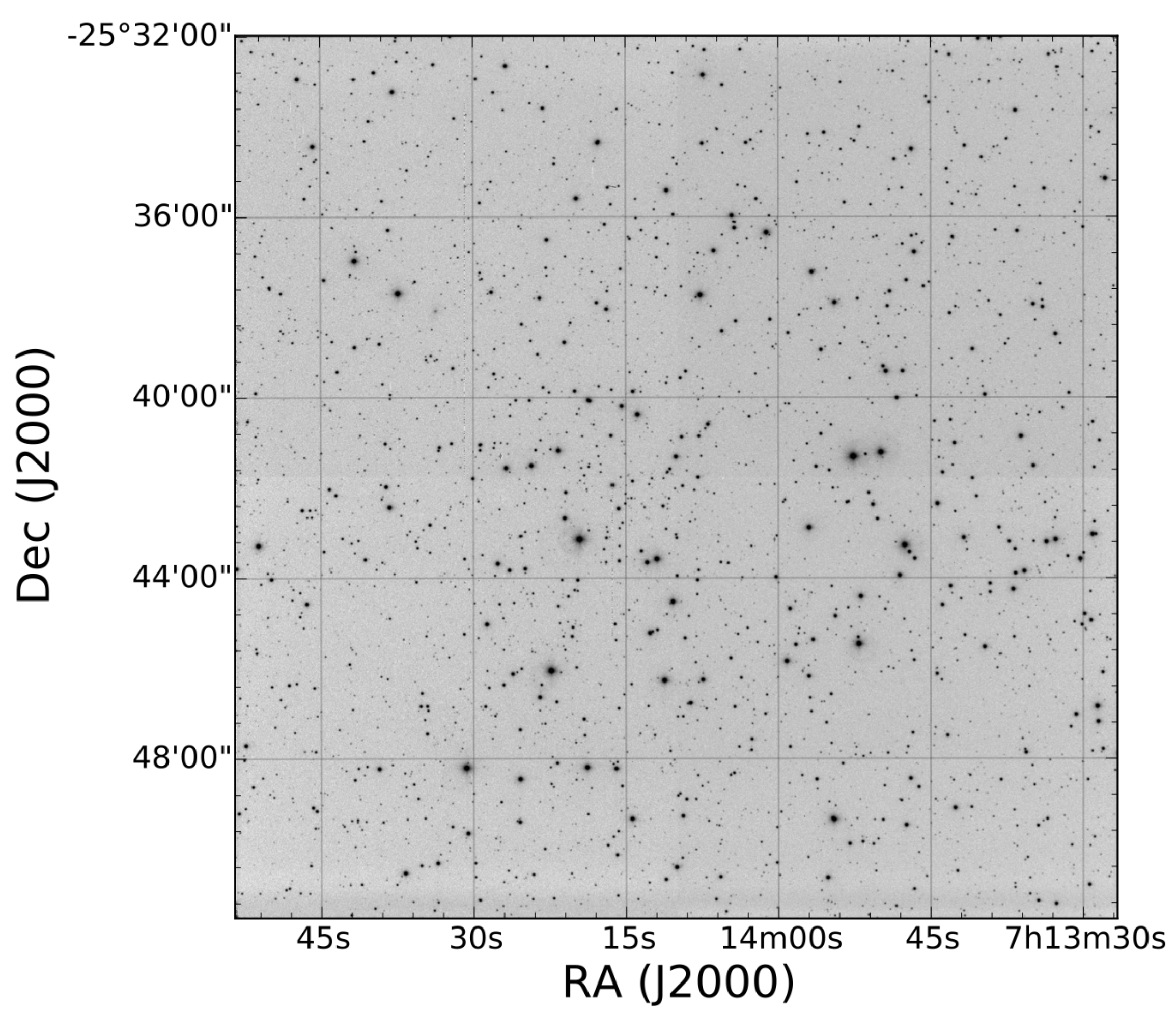}
  \includegraphics[scale=0.35]{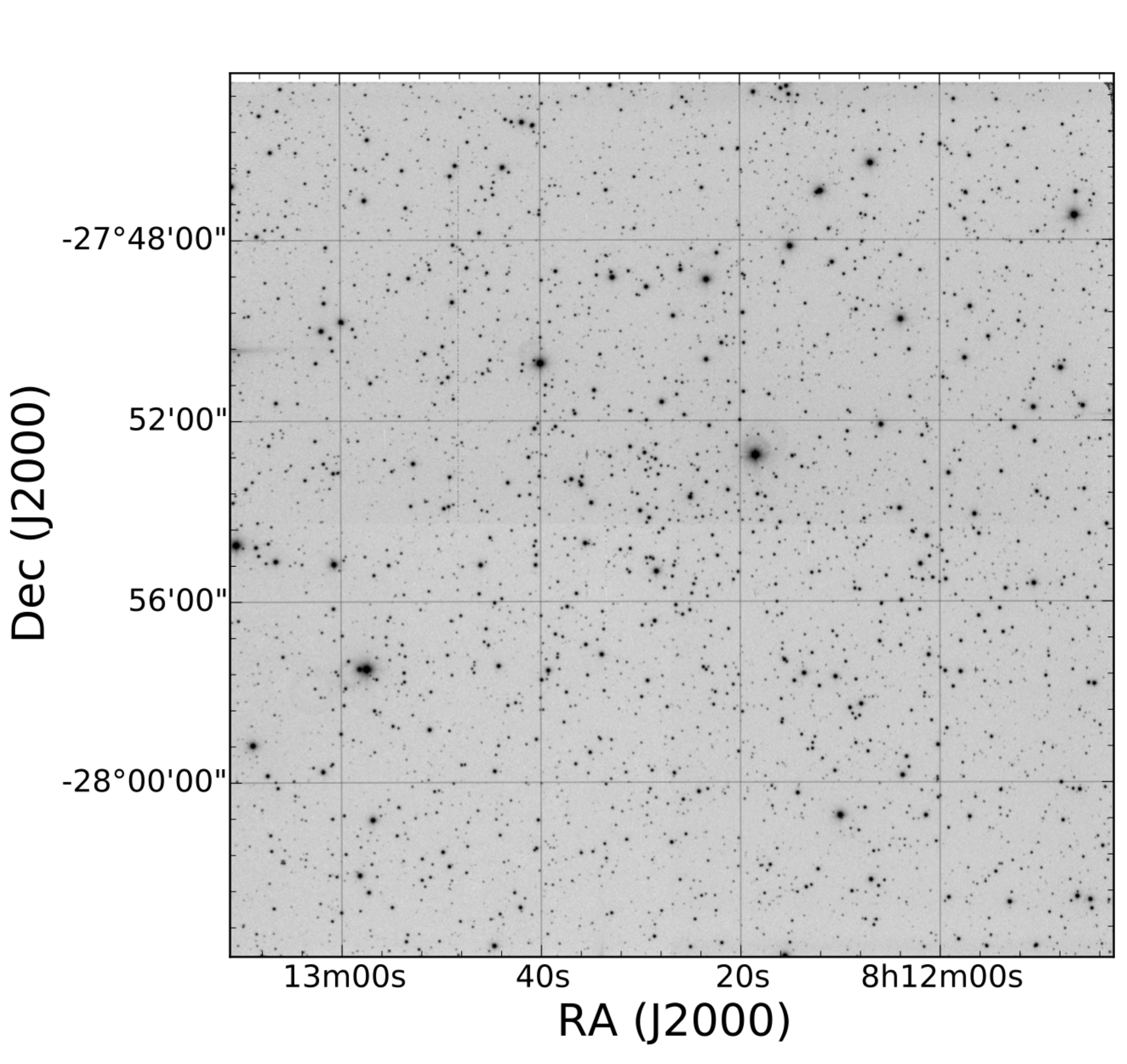}
  \includegraphics[scale=0.53]{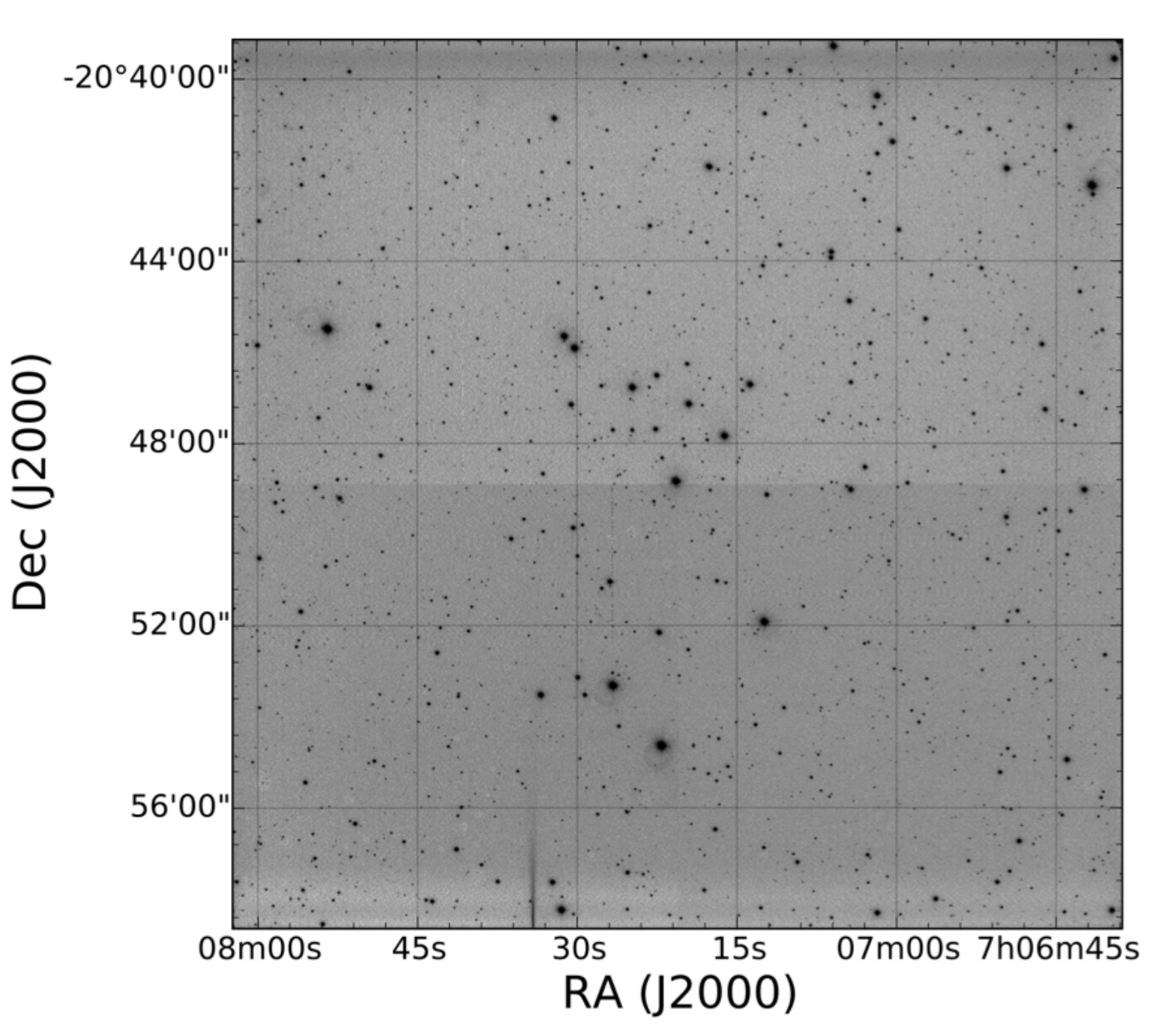}
  \includegraphics[scale=0.35]{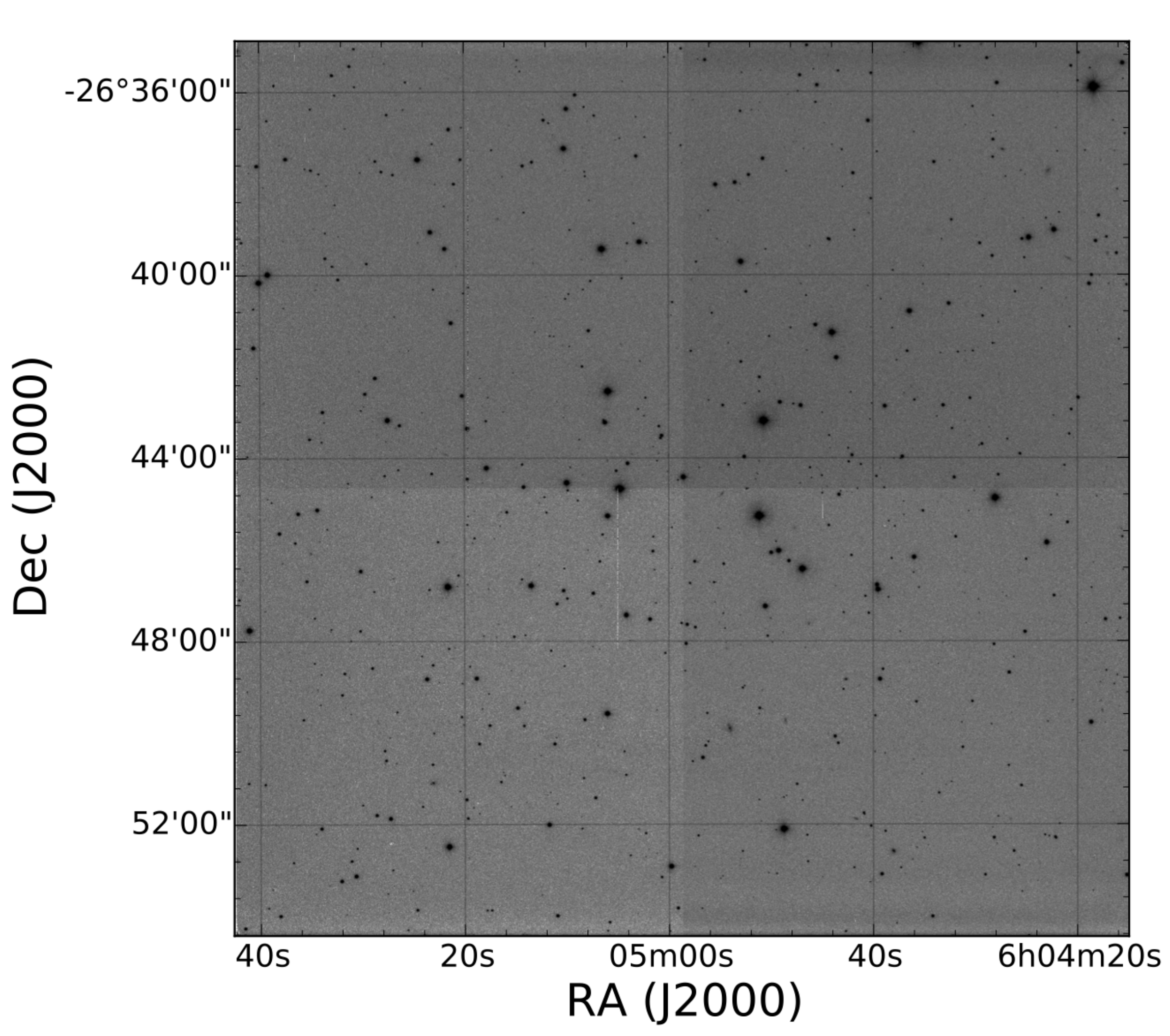}
  \caption{Bias and flat-field corrected images (150$^{\rm m}$ exposure) in B for the five program fields. 
  Bottom row: ESO489SC01. Middle row: Haffner 22 and Ruprecht~11. Top row: NGC~2215 and NGC~2354. Each field is 20 arcmin on a side.}
    \end{figure*}

 \begin{table}
\tabcolsep 0.1truecm
\caption{Equatorial and Galactic coordinates of program fields, as listed by WEBDA.}
\begin{tabular}{lcccc}
\hline
Cluster & RA(2000.0) & Dec (2000.0) &  $l$ & $b$ \\
\hline
   & [hh:mm:ss]  & [dd:mm:ss] & [deg] & [deg] \\
\hline\hline
ESO489SC01   &  06:04:58  &  -26:44:00 & 232.928  & -21.419 \\
NGC~2215       &  06:20:49  &  -07:17:00 & 215.993  & -10.102 \\
Ruprecht~11     &  07:07:21  &  -20:48:00 & 233.267  &  -5.981\\
NGC~2354       &  07:14:10  &  -25:41:24 & 238.368  &  -6.792\\
Haffner~22       &  08:12:27  &  -27:54:00 &  246.775  &   3.377\\
\hline\hline
\end{tabular}
\end{table}

\section{Literature overview}
The fields studied here contain catalogued Galactic open clusters that received  different levels of attention previously. Although this study focuses on fields peripheral to each cluster, their presence offers the possibility of assessing the quality of our photometry as well as deriving parameters for those clusters not studied previously. It is therefore important to review what is known about each.\\

\noindent
{\bf NGC~2215}\\
NGC~2215 was studied recently by Fitzgerald et al. (2015), who provide a summary of previous investigations with emphasis on the large differences in its parameters obtained over the years. It is not necessary to repeat their discussion here, but it is worth noting that large discrepancies in the derived parameters for clusters are unfortunately quite common in the literature (Netopil et al. 2015). The photometry of Fitzgerald et al. (2015) covers a region measuring $\sim$ 9 arcmin on a side, suitable for sampling the cluster core, which appears to lie within a radius of $\sim$5 arcmin. Their derived parameters are: distance $\sim$800 pc from the Sun, reddening E$_{B-V}=0.26$, and age slightly less than one Gyr. The cluster seems to harbour only one red giant, for which Mermilliod et al. (2008) measured a radial velocity.  Its membership is difficult to assess, since there are no radial velocity measures for main sequence stars.

Two additional points render the cluster particularly interesting. First, the metallicity found by Fitzgerald et al. (2015) is very low for the age and Galactic location of the cluster (Magrini et al. 2009). Second, the authors refer to possible contamination from Perseus arm stars lying background to the cluster, a feature that could not be tested properly given their small field of view. Such a population would be as old as the Sun ($\sim$ 4.5 Gyr), but significantly more metal poor ([Fe/H] $\sim$ -0.8), and located at a heliocentric distance of $\sim$ 4 kpc.\\
These parameters are, however,  only suggested, and not inferred via  isochrone fitting or any other method, and no clear mention is done of which stars would constitute this old population. 
This goes without saying that  a 4.5 Gyr population would hardly be part of a spiral arm, which is expected to be
mostly composed of gas, dust and very young stars.\\

\noindent
{\bf NGC~2354}\\
This cluster has not been the subject of a global study previously, the most comprehensive study being that of Claria et al. (1999), although limited to the red giants. Those authors note that {\it A better age determination is awaiting more precise CCD data to replace the old, photographic data of Durbeck (1960) and the determination of membership of main-sequence stars}. Their solution, based on available data, is: reddening E$_{B-V}=0.13$, age $\sim$1 Gyr, and distance about 1.5 kpc. The results largely differ from more recent estimates obtained by Kharchenko et al. (2005, 2013), who suggest a much younger age (200 Myr) and a distance of almost 4 kpc. A recent spectroscopy study by Reddy et al. (2015) finds the cluster to be metal poorer than the Sun ($[Fe/H=-0.19\pm0.04]$)\\

\noindent
{\bf Haffner~22}\\
This cluster is unstudied, although from available data Kharckenko et al. (2013) derived an age of 1.54 Gyr, a distance of 2796 pc, and a reddening E$_{B-V}=0.208$.\\

\noindent
{\bf Ruprecht~11}\\
No formal study of this cluster is available in the literature. From available data Kharckenko et al. (2013) derived an age of 630 Myr, a distance of 1958 pc, and a reddening E$_{B-V}=0.437$.\\

\noindent
{\bf ESO489~SC01}\\
To our knowledge no study of this cluster is available in the literature.

\begin{table}
\tabcolsep 0.02truecm
\caption{A log of {\it UBVI} photometric observations for the fields of study.}
\begin{tabular}{lcccc}
\hline
\noalign{\smallskip}
Date & Field & Filter & Exposures (sec) & airmass (X)\\
\noalign{\smallskip}
\hline\hline
\noalign{\smallskip}
2006 Mar 19    & Ruprecht 11 & $U$ & 2x30, 200, 1800 & 1.04$-$1.09\\
                     &                        & $B$ & 30,150,900 & 1.01$-$1.02\\
                      &                       & $V$ & 10, 30,100, 700 & 1.02$-$1.10 \\
                       &                      & $I$  & 5, 10, 30,100,600 & 1.02$-$1.11 \\
2009 Mar 18 & ESO489SC01 & $U$ & 30,200,2000 & 1.08$-$1.09\\
                        &                         & $B$ &20,150,1500& 1.25$-$1.27\\
                         &                         & $V$& 10,100,900& 1.17$-$1.19\\
                          &                        & $I$& 10,100,900& 1.39$-$1.42\\ 
2009 Mar 17 & NGC 2354 & $V$ & 3x100,900   & 1.07$-$1.08 \\
            &            & $B$ & 20,150, 1500     & 1.18$-$1.20 \\
            &            & $U$ & 30, 200,2000       & 1.02\\
            &            & $I$ & 2x100,900   & 1.12$-$1.13 \\
2009 Mar 19 & NGC 2215 & $U$ & 30,2x200,2000     & 1.17$-$1.19 \\
            &            & $B$ & 2x20, 150,1500      & 1.30$-$1.33 \\
                      & Haffner 22   &  $U$ & 30,200,2000 & 1.29$-$1.32\\
                        &                     &  $B$ & 2x20, 150, 1500 & 1.05$-$1.06\\                      
2009  Mar 20 & NGC 2215  & $V$ & 3x10,900     & 1.16$-$1.17 \\
            &            & $I$ & 10,100,900   & 1.21$-$1.23 \\
               & Haffner 22 & $V$ & 10,100,900 & 1.02 \\
                   &     & $I$ & 10.100.900 & 1.01\\                   
2009  Mar 21 &  NGC 2354  & $V$ & 20,100,900   & 1.15$-$1.16 \\
            &            & $B$ & 20,150, 1500     & 1.05$-$1.06 \\
            &            & $U$ & 200,2000       & 1.10$-$1.02\\
            &            & $I$ & 20, 100,900   & 1.10$-$1.11 \\          
\hline
\hline
\end{tabular}
\end{table}

\begin{table*}
\tabcolsep 0.4truecm
\caption{Night by night photometric solutions.}
\begin{tabular}{cccccc}
\hline\hline
Night  &    19/03/2006  &    18/03/2009 &     19/03/2009  &        20/03/2009  &        21/03/2009 \\
\hline\hline
u1    & -0.747$\pm$0.003   &-0.884$\pm$0.006& -0.860 $\pm$0.008  &  -0.879$\pm$0.007  & -0.885$\pm$0.007\\
u2    &  0.45                        & &                                 &                                  &\\
u3    & -0.015$\pm$0.006   &-0.037$\pm$0.009&-0.017 $\pm$0.012   & -0.023$\pm$0.010   & -0.016$\pm$0.010\\
rms  &  0.02                        &0.09 &   0.10                       &  0.09                         &     0.08\\
\hline
b1             &-1.980$\pm$0.013    &-2.085$\pm$0.010 &   -2.063$\pm$0.010 &   -2.068$\pm$0.010 &   -2.081$\pm$0.010\\
b2             & 0.25                         &                               &                               &                                  &        \\
b3             & 0.153 $\pm$0.017    &0.150$\pm$0.010 & 0.128$\pm$0.010    &  0.132$\pm$0.010   &    0.132$\pm$0.010\\
rms           & 0.03                          &       0.08                & 0.05                       &         0.08                 &              0.08\\
\hline
v1$_{bv}$  & -2.199$\pm$0.009   &-2.136$\pm$0.028   &   -2.126$\pm$0.006 &   -2.127$\pm$0.006  &  -2.139$\pm$0.007\\
v2$_{bv}$  &  0.16                         &                                &                                 &                                  & \\
v3$_{bv}$  &-0.060$\pm$0.012    & -0.028$\pm$0.005 &    -0.035$\pm$0.006 &    -0.036$\pm$0.006 &    -0.021$\pm$0.006\\
rms            &  0.02                        &      0.04                     & 0.05                           &    0.05                       &    0.05\\
\hline
v1$_{vi}$ & -2.198$\pm$ 0.005 &-2.148$\pm$0.005&    -2.121$\pm$0.006 &    -2.124$\pm$ 0.006  &   -2.159 $\pm$0.007\\
v2$_{vi}$  & 0.16                     & & &\\
v3$_{vi}$ &-0.063$\pm$0.005 &-0.013$\pm$0.045&   -0.038$\pm$0.005 &    -0.032$\pm$0.005 &     0.001 $\pm$ 0.005\\
rms   & 0.03                             &  0.05&   0.05       &        0.04        &       0.05\\
\hline
i1   &  -1.298$\pm$0.009  &  -1.321$\pm$0.004& -1.319$\pm$0.005  &  -1.313$\pm$0.005  &  -1.316$\pm$0.005\\
i2    &   0.08                      &                                &  &                                 &                              \\
i3    & -0.054$\pm$0.009  & -0.010$\pm$0.003& -0.017$\pm$0.004   & -0.014$\pm$0.003    &-0.016$\pm$0.004\\
rms &   0.02                      & 0.04 & 0.04                       &   0.04                        &  0.04                      \\
\hline\hline
\end{tabular}
\end{table*}

  \begin{figure}
  \includegraphics[width=\columnwidth]{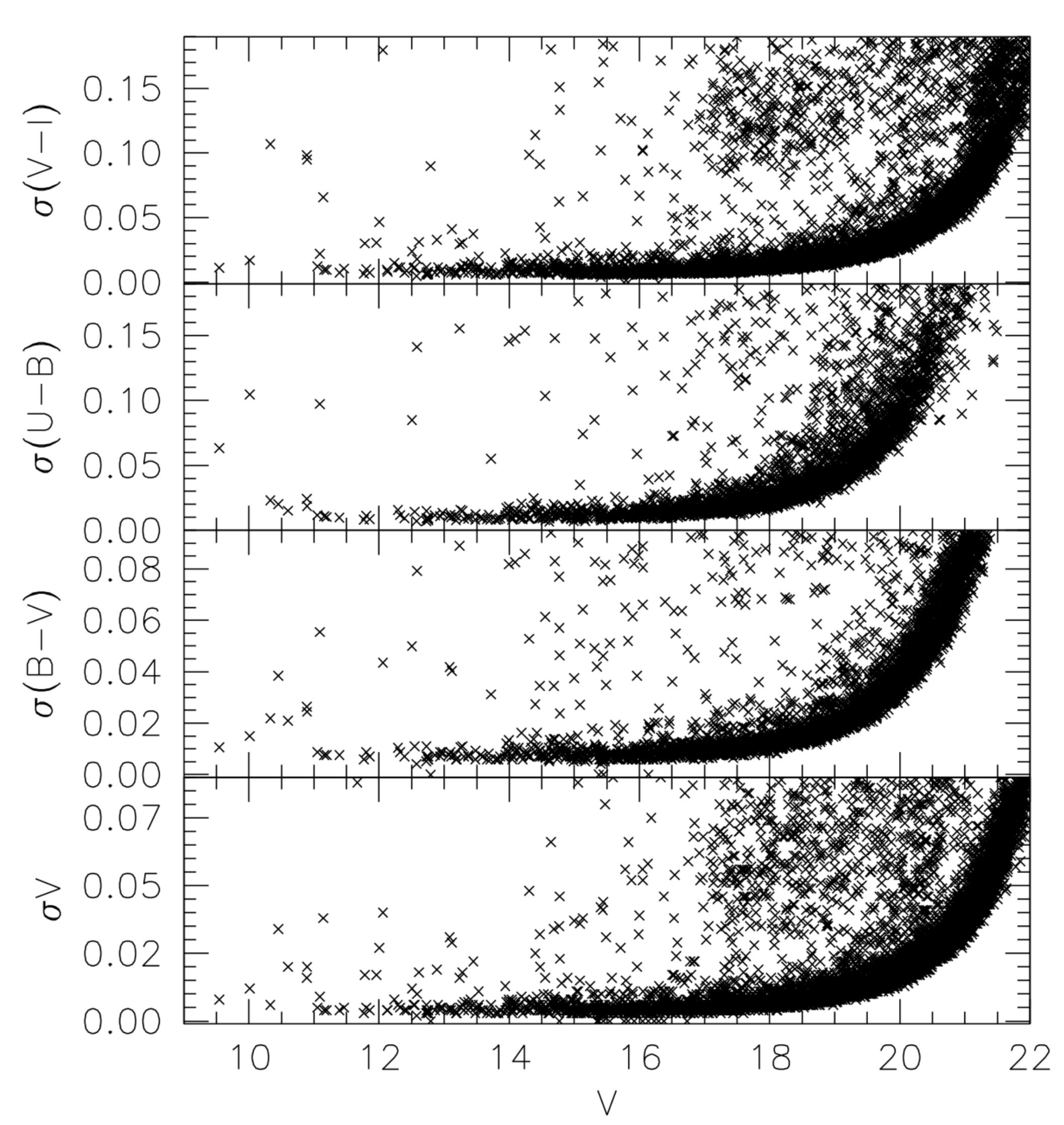}
   \caption{The trend of global photometric uncertainties as a function of {\it V} magnitude for the NGC~2354 field.}
    \end{figure}

 \begin{figure}
  \includegraphics[width=\columnwidth]{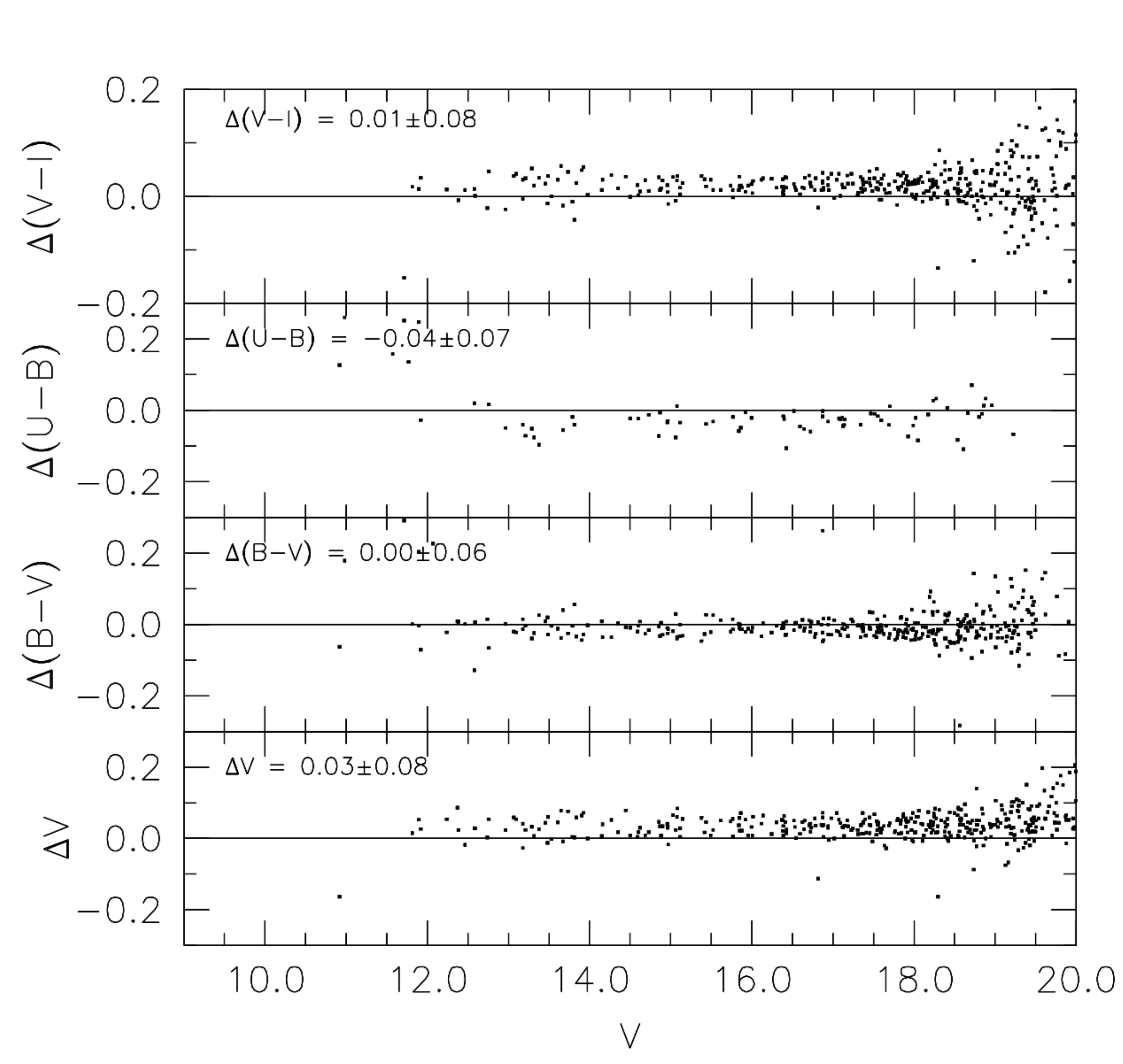}
   \caption{A comparison of our photometry with that of Fitzgerald et al. (2015).}
    \end{figure}

\section{Observations and data reduction}
Observations of program clusters were obtained in March 2006 and 2009 using the Y4KCAM camera attached to the Cerro Tololo Inter-American Observatory (CTIO, Chile) 1-m telescope, operated by the SMARTS consortium\footnote{http://www.astro.yale.edu/smarts}. The camera is equipped with an STA~$4064\times4064$ CCD\footnote{http://www.astronomy.ohio-state.edu/Y4KCam/detector.html} with 15-$\mu$m pixels, yielding a scale of 0$^{\prime\prime}$.289 pixel$^{-1}$ and a field-of-view (FOV) of $20^{\prime}\times 20^{\prime}$ at the Cassegrain focus of the telescope.

The FOV is large enough to incorporate each cluster's core region (see below), and also sample the surrounding Galactic field. The pointings are shown in Fig.~1, which contains CCD images for the five fields.

Table~2 presents a log of our {\it UBVI} observations, all carried out in photometric, good-seeing conditions. Our instrumental {\it UBVI} photometric system was defined by use of a standard broad-band Kitt Peak {\it UBVI}$_{kc}$ set of filters\footnote{http://www.astronomy.ohio-state.edu/Y4KCam/filters.html}.
Transformation from our instrumental system to the standard Johnson-Kron-Cousins system, and corrections for extinction, were established each night by observing Landolt's area PG~1047 and SA~98 (Landolt 1992) multiple times, through air-masses as close as possible to those for the target clusters. Field SA~98 in particular includes over 40 well-observed standard stars, with good magnitude and color coverage: $9.5\leq V\leq15.8$, $-0.2\leq(B-V)\leq2.2$, $-0.3\leq(U-B)\leq2.1$.

The basic calibration of the CCD frames was done using the Yale/SMARTS {\it Y4K} reduction script based on the IRAF\footnote{IRAF is distributed by the National Optical Astronomy Observatory, which is operated by the Association of Universities for Research in Astronomy, Inc., under cooperative agreement with the National Science Foundation.} package \textsc{ccdred}, and the photometry was performed using IRAF's \textsc{daophot} and \textsc{photcal} packages. Instrumental magnitudes were extracted following the point spread function (PSF) method (Stetson 1987) using a quadratic, spatially-variable master PSF (PENNY function). Finally, the PSF photometry was aperture corrected using aperture
corrections determined making aperture photometry of bright, isolated stars in the field.

Aperture photometry was then carried out for all stars using the PHOTCAL package, with transformation equations of the form:\\

\begin{equation}
u = U + u1 + u3 \times(U--B) + u2 \times X               
\end{equation}

\begin{equation}    
b = B + b1 + b3 \times(B--V) + b2 \times X  
\end{equation}

\begin{equation}                
v = V + v1_{bv} + v3_{bv} \times (B--V) + v2_{bv} \times X  
\end{equation} 

\begin{equation}
v = V + v1_{vi} + v3_{vi} \times (V--I) + v2_{vi} \times X  
\end{equation}

\begin{equation} 
i = I_C + i1 + i3 \times (V--I) + i2 \times X                
\end{equation}

\noindent
where {\it UBVI$_C$} and {\it ubvi} are standard and instrumental magnitudes, respectively, and {\it X} is the air mass of the observation. Typical values for the extinction coefficients for CTIO (see Baume et al. 2011) were adopted. {\it V} magnitudes were derived using Eq. 3 when the {\it B} magnitude was available, otherwise Eq.~4. The calibration coefficients and their uncertainties are shown in Table~3.

The trend of global (PSF plus calibration) photometric uncertainties is shown in Fig.~2. Note that they are well below 0.05 mag to $V\sim19.5$ for all color combinations.
  
World Coordinate System (WCS) header information for each frame was obtained using the ALADIN tool and 2MASS data (Skrutskie et al. 2006). The procedure to perform astrometric calibration of our data is explained by Baume et al. (2009), and used here to obtain a reliable astrometric calibration ($\sim0^{\prime\prime}.12$).

The STILTS (Taylor 2006) tool was used to manipulate tables and to cross-correlate the {\it UBVI}$_C$ and {\it JHK} 2MASS data. In the case of Ruprecht~11 and NGC~2354, a few bright stars were saturated in our images, therefore we complemented our catalogues with photometric data from APASS (Henden et al. 2010). The result was a catalogue with astrometric/photometric information for detected objects covering a FOV of approximately $20^{\prime}\times 20^{\prime}$ for each cluster (as in Fig.~1). The full catalogues are made available in electronic form at the CDS website.

Finally, our photometric data for NGC~2215 were compared with data in the recent study by Fitzgerald et al. (2015) in order to assess the quality and zero-points of our observations. From a grand-total of 498 stars in common we find:\\

\noindent
$\Delta$ V       = 0.03$\pm$0.08,\\

\noindent
$\Delta$ (B--V) = -0.00$\pm$0.06,\\

\noindent
$\Delta$ (U--B) = -0.04$\pm$0.07, and\\

\noindent
$\Delta$ (V--I)   = 0.01$\pm$0.08.\\

\noindent
in the sense this study minus Fitzgerald et al. The comparison is shown in Fig.~3, where one can see that the scatter is quite large for stars fainter than V$\sim$18, a result of shallower limits for the Fitzgerald et al. (2015) study. If the comparison is restricted to stars in the range $12 \leq V \leq 17$, the differences read:\\

\noindent
$\Delta$ V       = 0.02$\pm$0.03,\\

\noindent
$\Delta$ (B--V) = -0.00$\pm$0.03,\\

\noindent
$\Delta$ (U--B) = 0.03$\pm$0.04, and\\

\noindent
$\Delta$ (V--I)   = 0.01$\pm$0.03.\\

\noindent 
There is generally good agreement between our observations and those of Fitzgerald et al. (2015), with no indication that our photometry deviates significantly from the standard system.
  
 \begin{figure*}
  \includegraphics[scale=0.35]{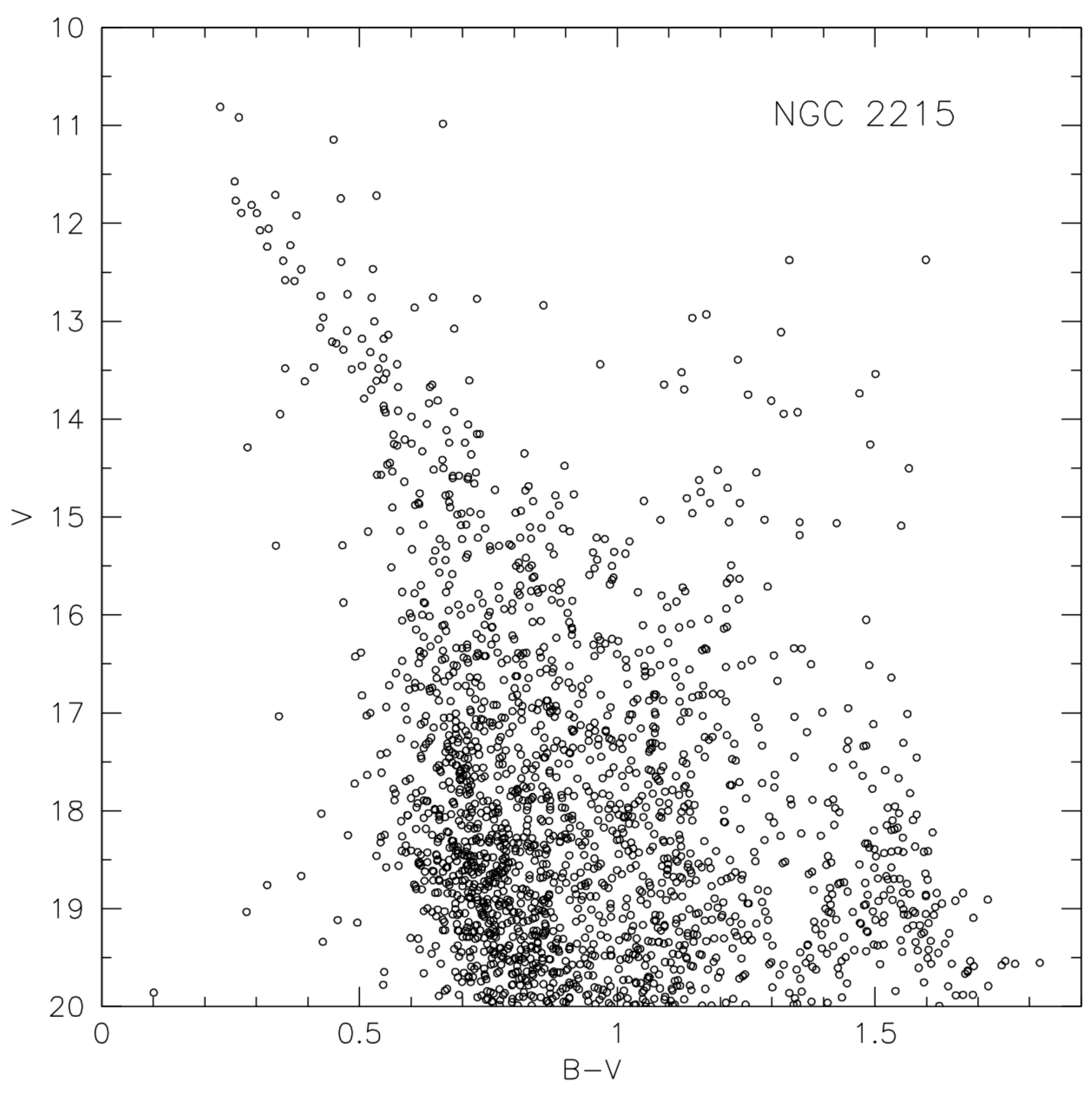}
   \includegraphics[scale=0.35]{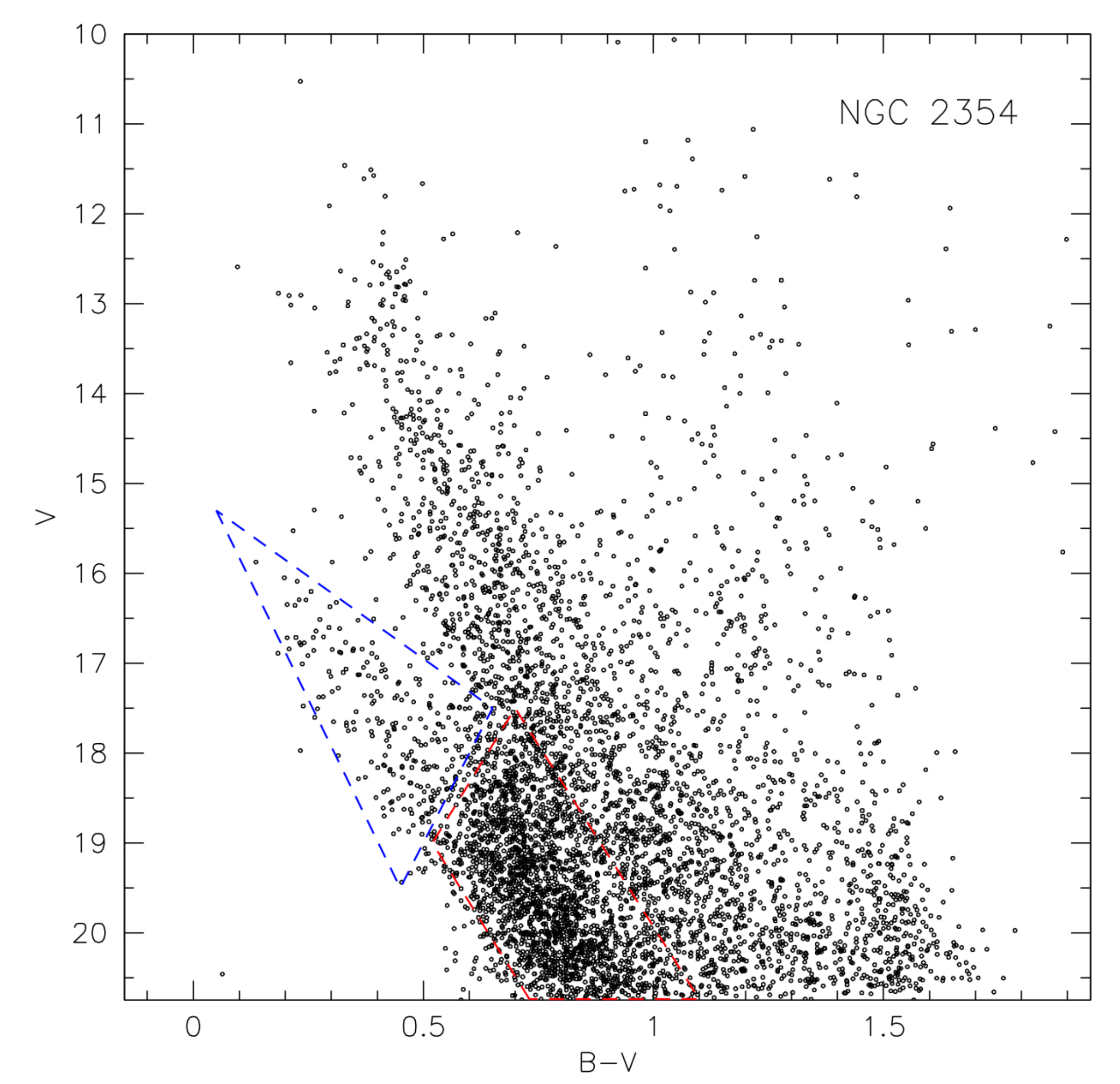}
  \includegraphics[scale=0.35]{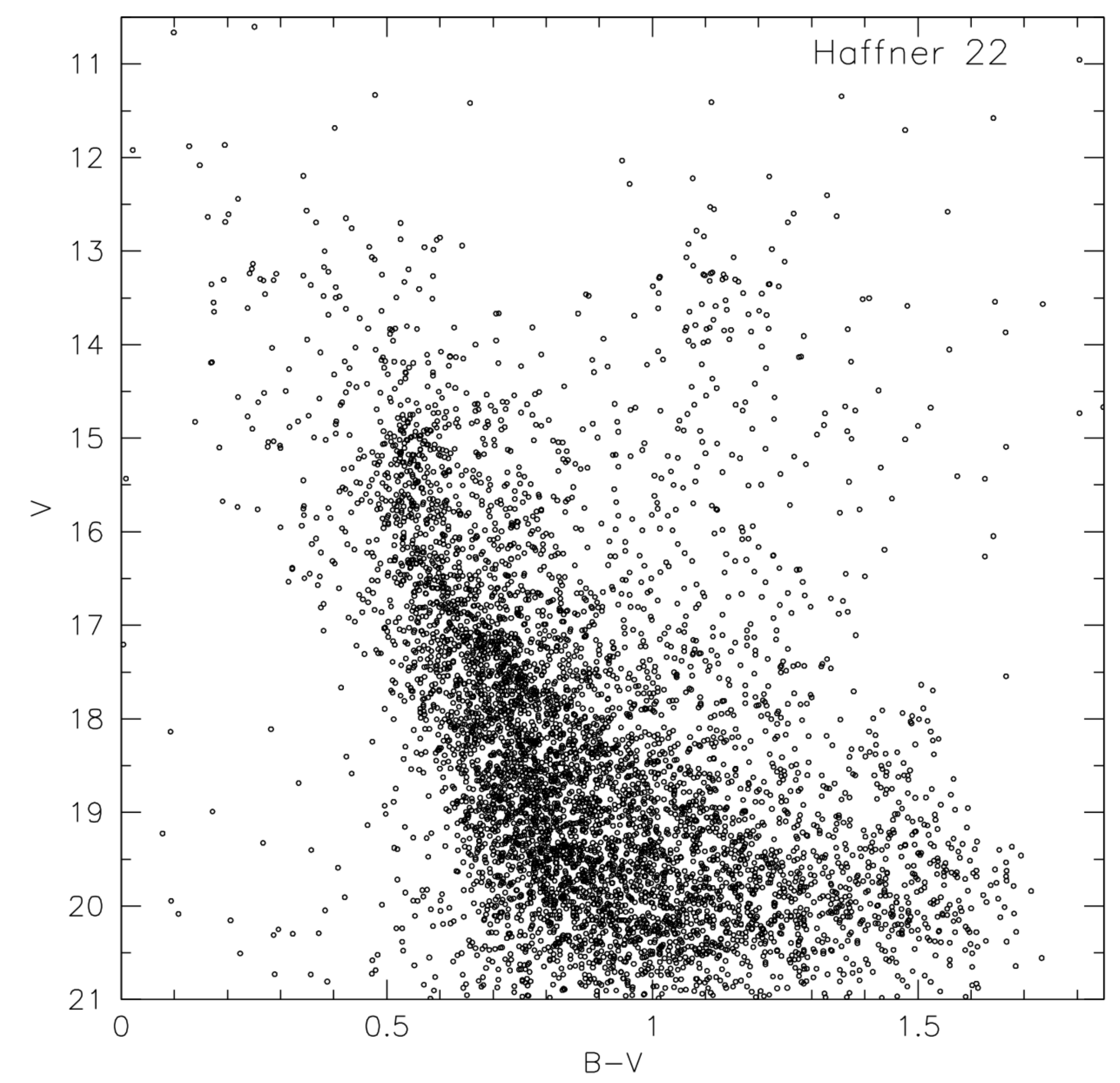}
  \includegraphics[scale=0.35]{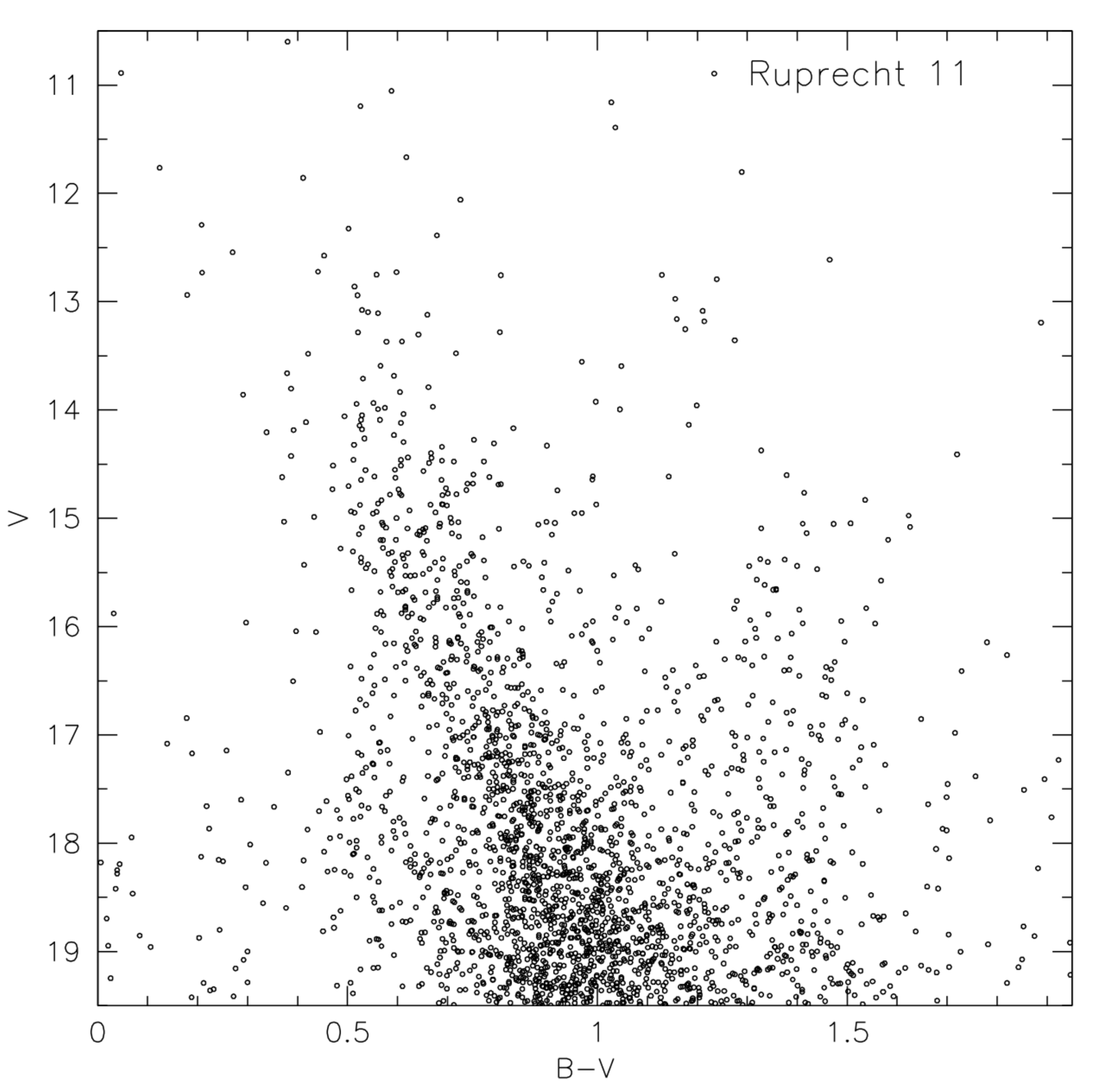}
  \includegraphics[scale=0.35]{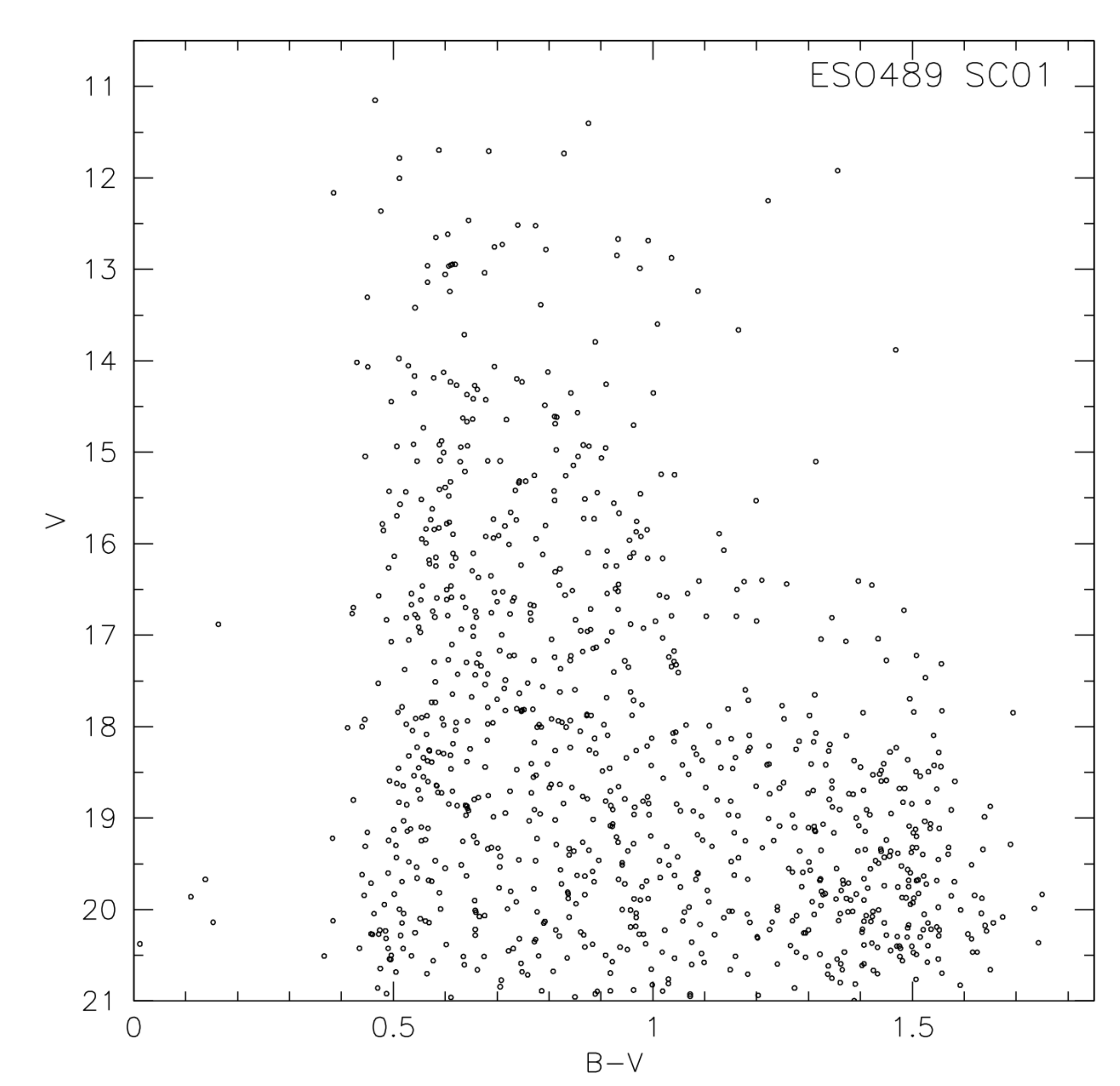}
  \caption{V/B-V CMD for the five program fields. 
  Bottom: ESO489SC01. Middle row: Haffner 22 and Ruprecht~11. Top row: NGC~2215 and NGC~2354. In NGC 2354 CMD we highlight the approximate location of the two most important features in the field. The dashed blue triangle comprises with blue the blue plume, young stars in the cluster background. The dashed red poligon contains the intermediate-age population.}
    \end{figure*}

\section{Color Magnitude Diagrams}
To set the scene, we begin by describing and discussing the general appearance of the color-magnitude diagrams (CMDs) for the five stellar fields. They are shown in Fig.~4, which contains all stars in each field with photometric uncertainties $\sigma_V\leq 0.03$ magnitude (see previous Section). A synoptic, global examination of the diagrams reveals the following features:\\

  \begin{figure*}
  \includegraphics[scale=0.19]{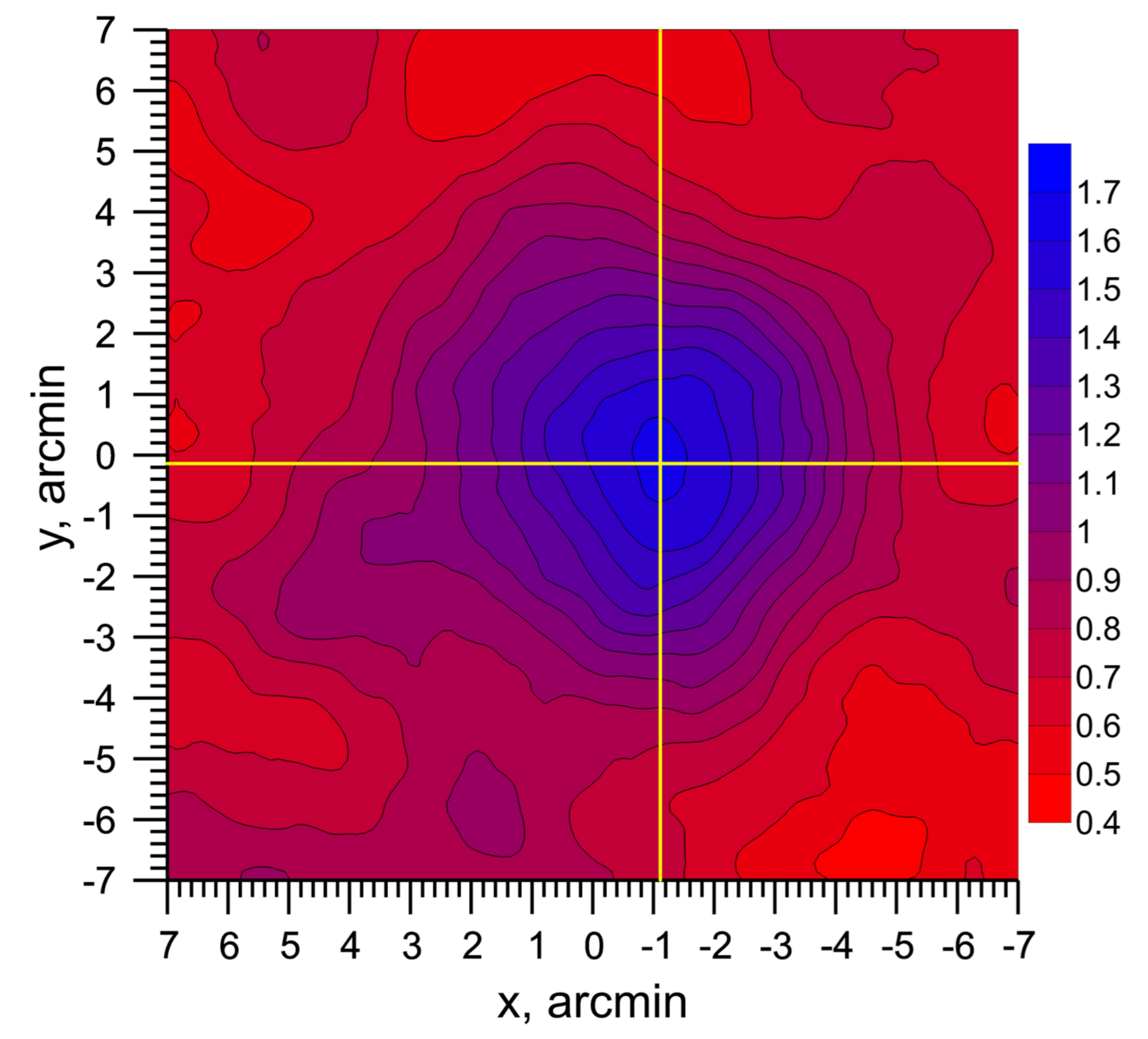}
   \includegraphics[scale=0.2]{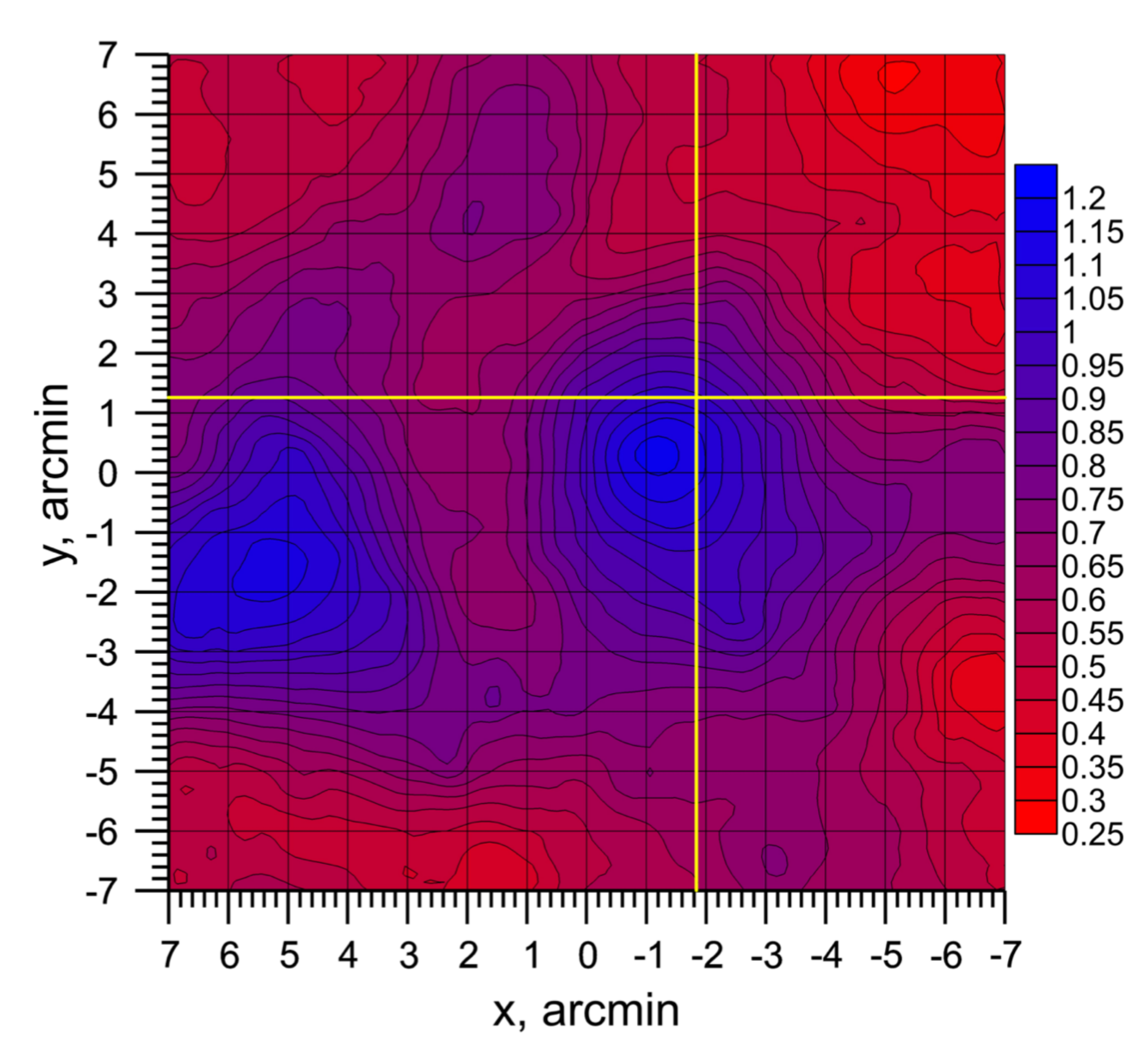}
     \includegraphics[scale=0.2]{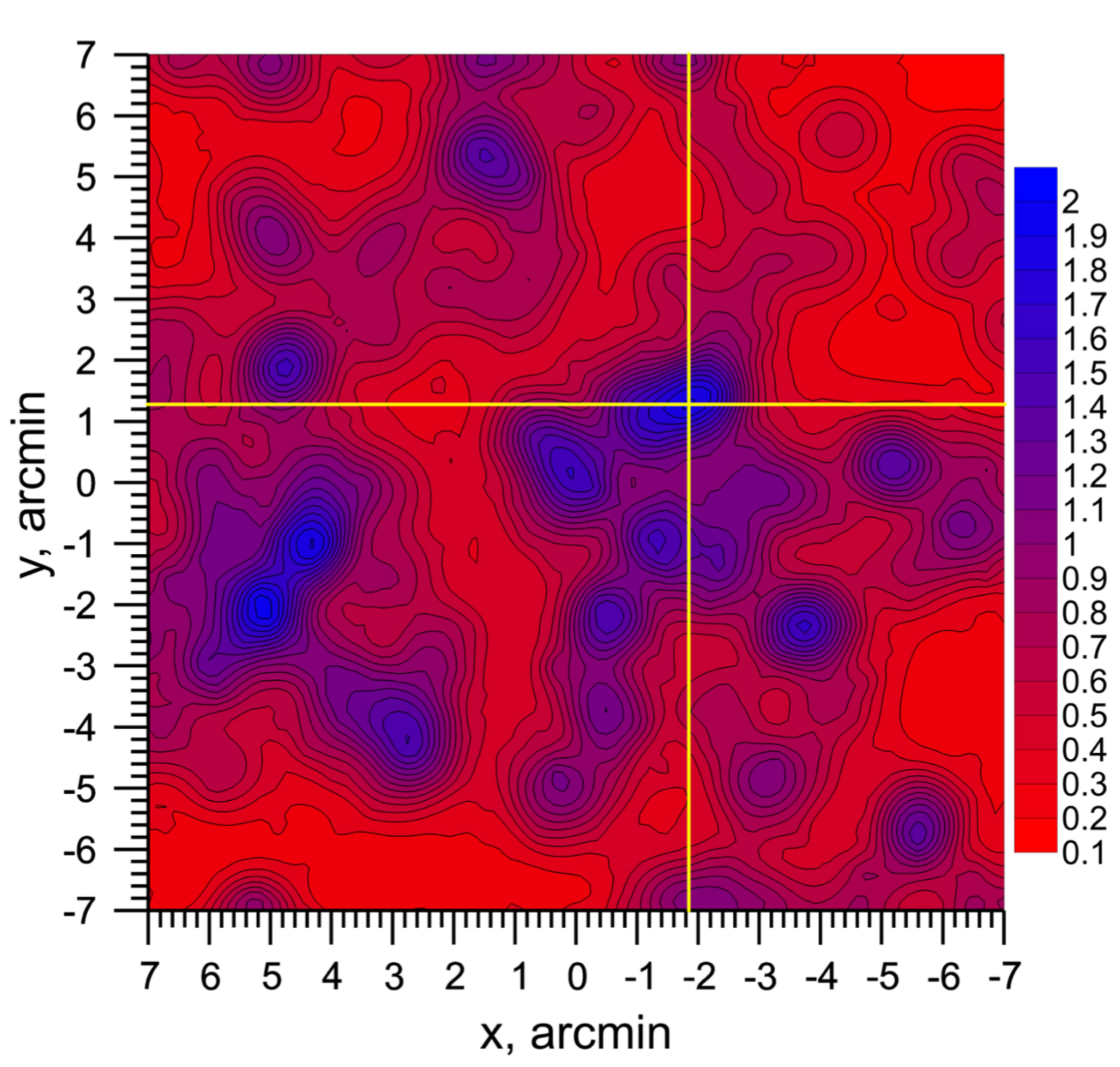}
  \includegraphics[scale=0.2]{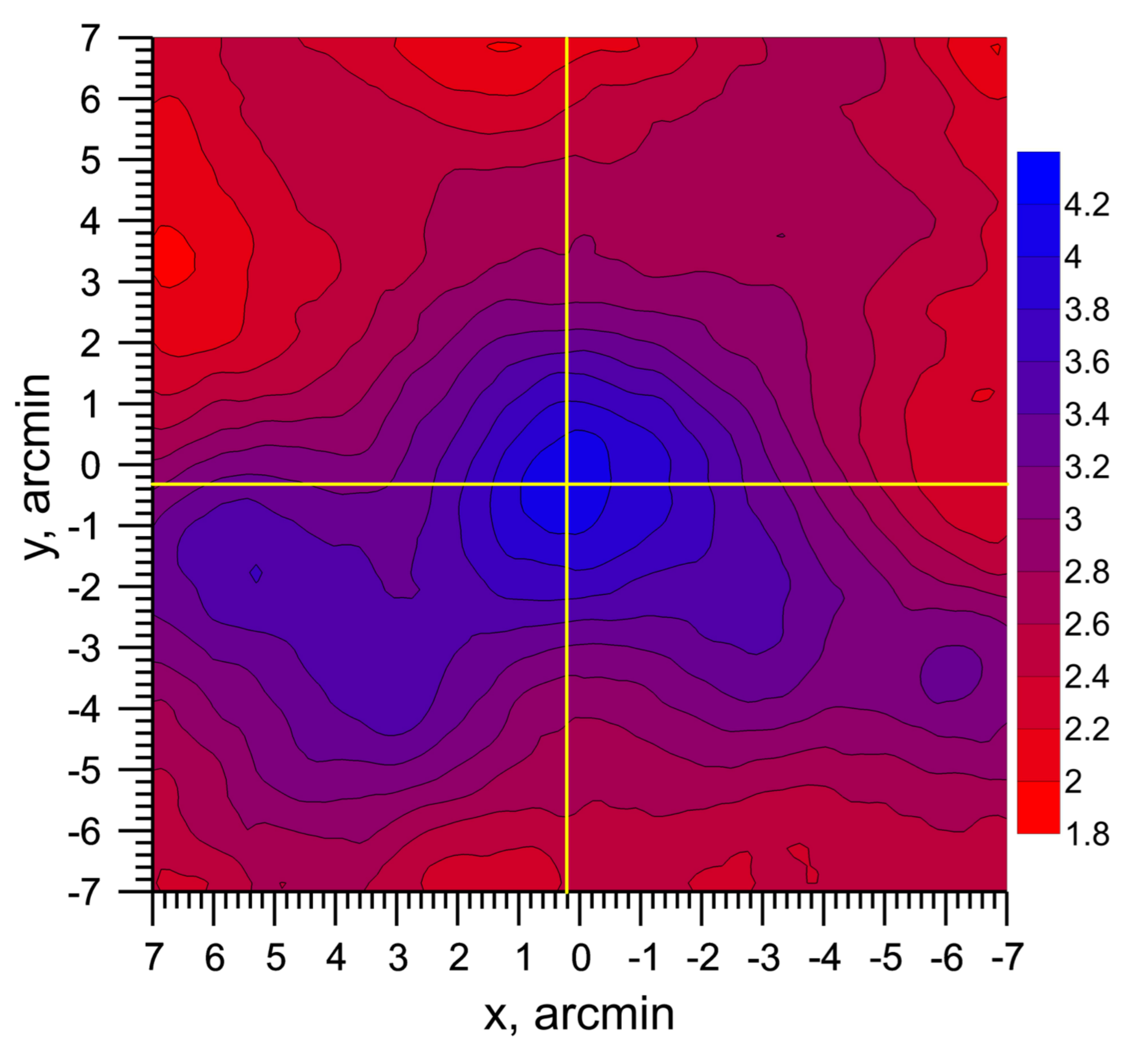}
  \includegraphics[scale=0.2]{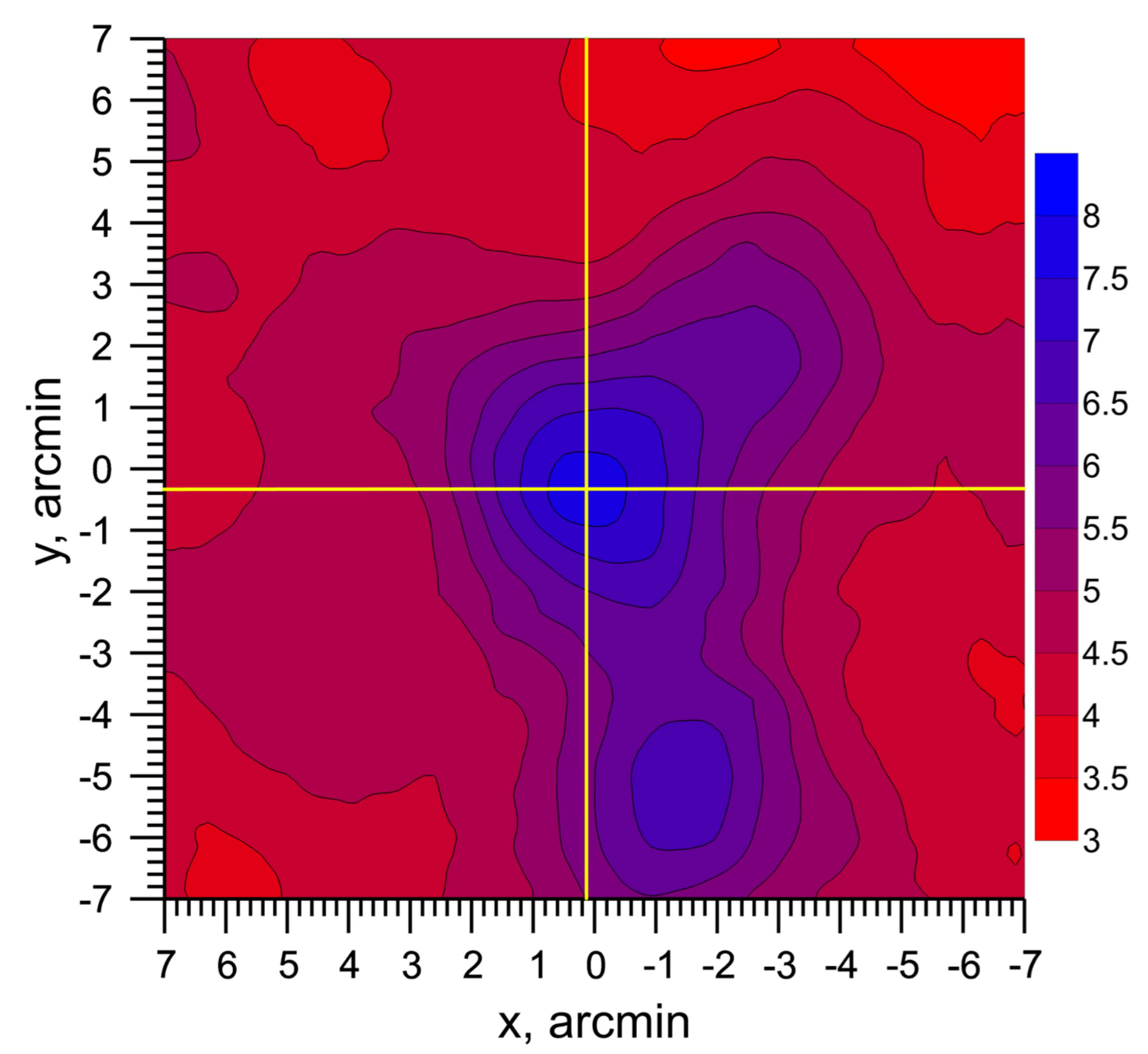}
  \includegraphics[scale=0.2]{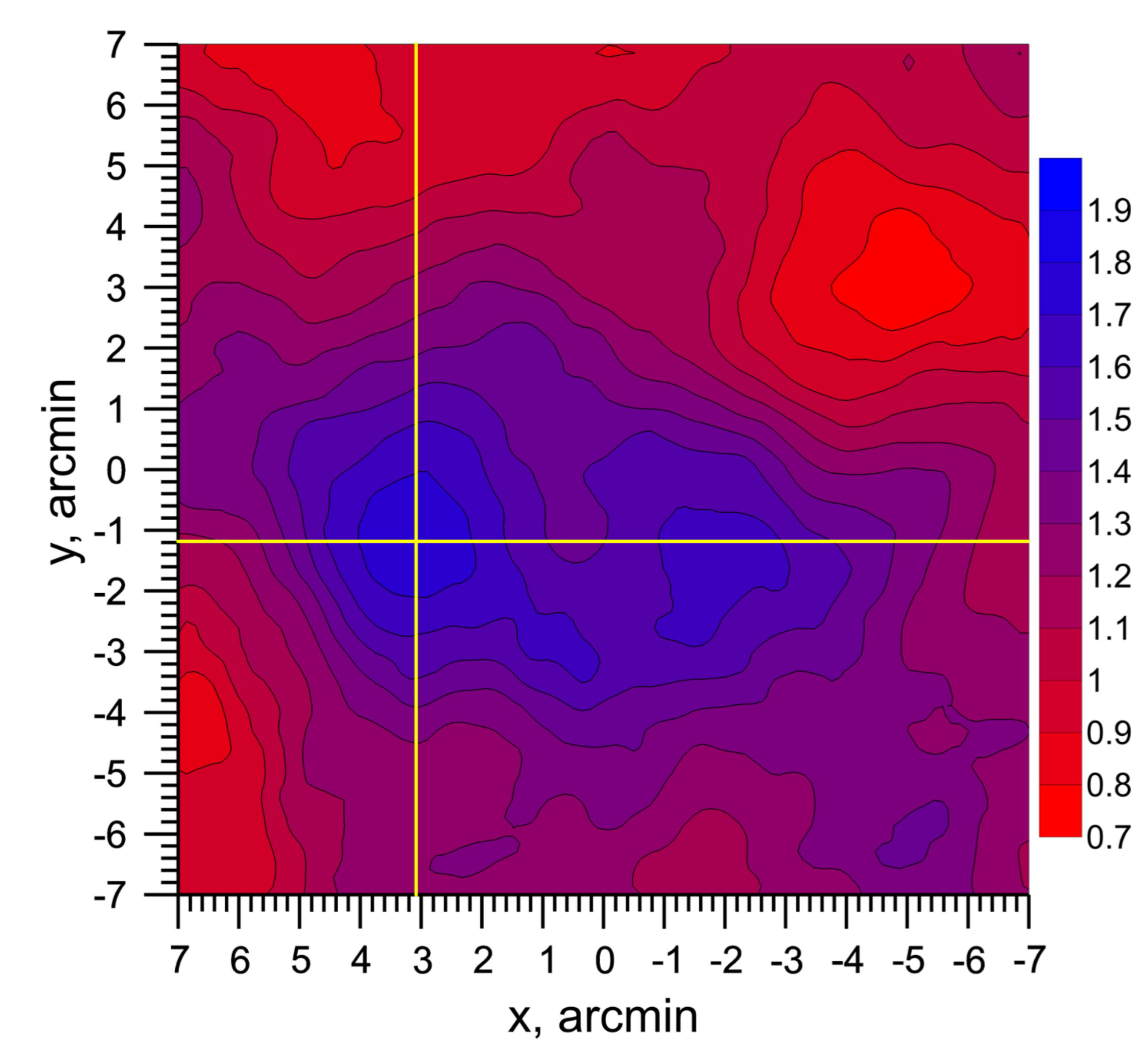}
  \caption{Surface density maps. Bottom: Haffner~22, Ruprecht 11, and ESO489SC01. {\bf Top, from the left to the right: NGC~2215, NGC~2354 (half kernel width value = 3 arcmin), and NGC~2354 (half-kernel width value = 1 arcmin). The yellow lines indicate the position of the derived cluster center. The density scale, color coded as in vertical bars, is in stars/arcmin2. All maps are plotted using 3 arcmin for the half-width kernel value. See text for additional details}.}
      \end{figure*}
    
\begin{figure*}
  \includegraphics[scale=0.25]{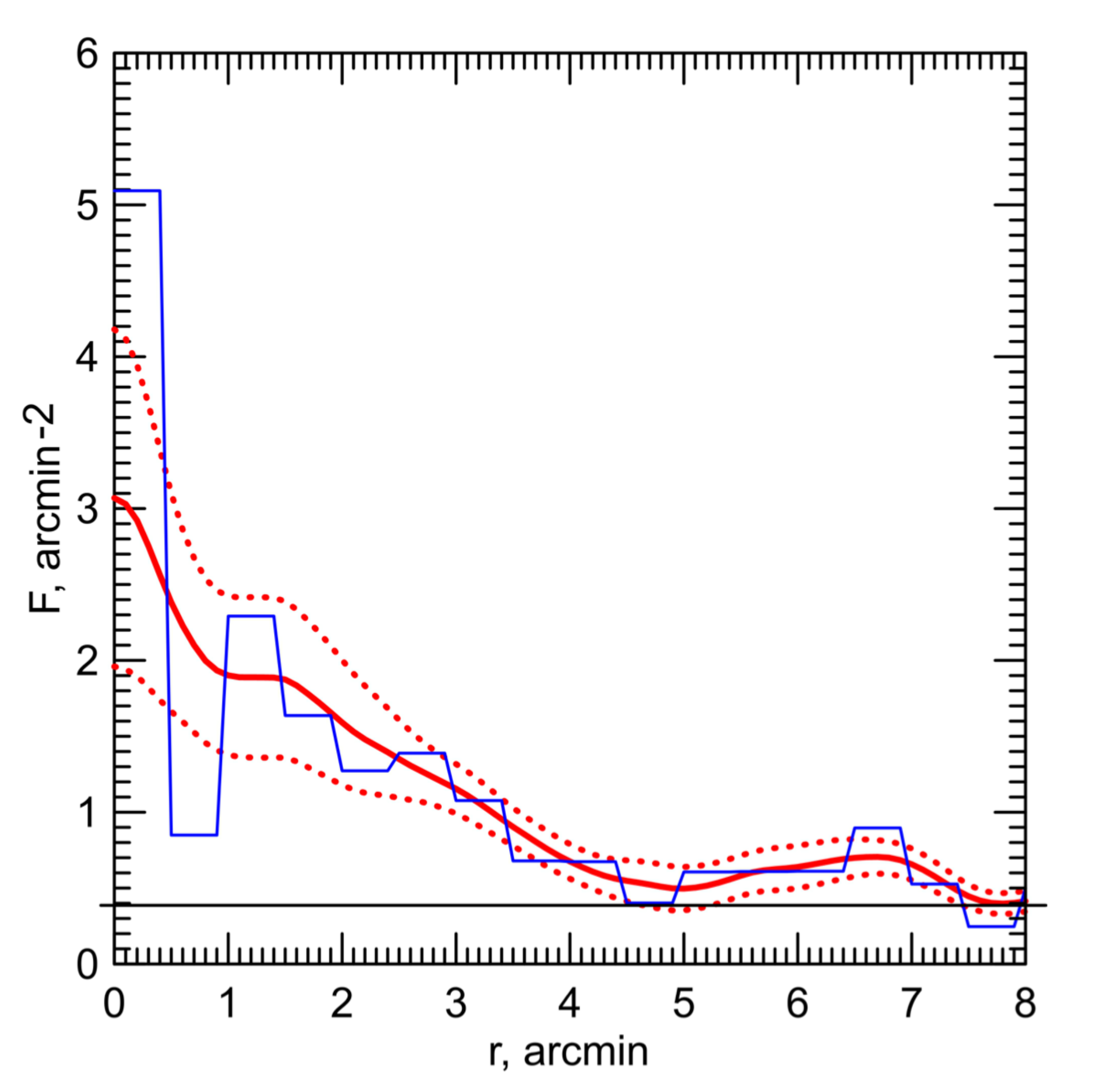}
   \includegraphics[scale=0.285]{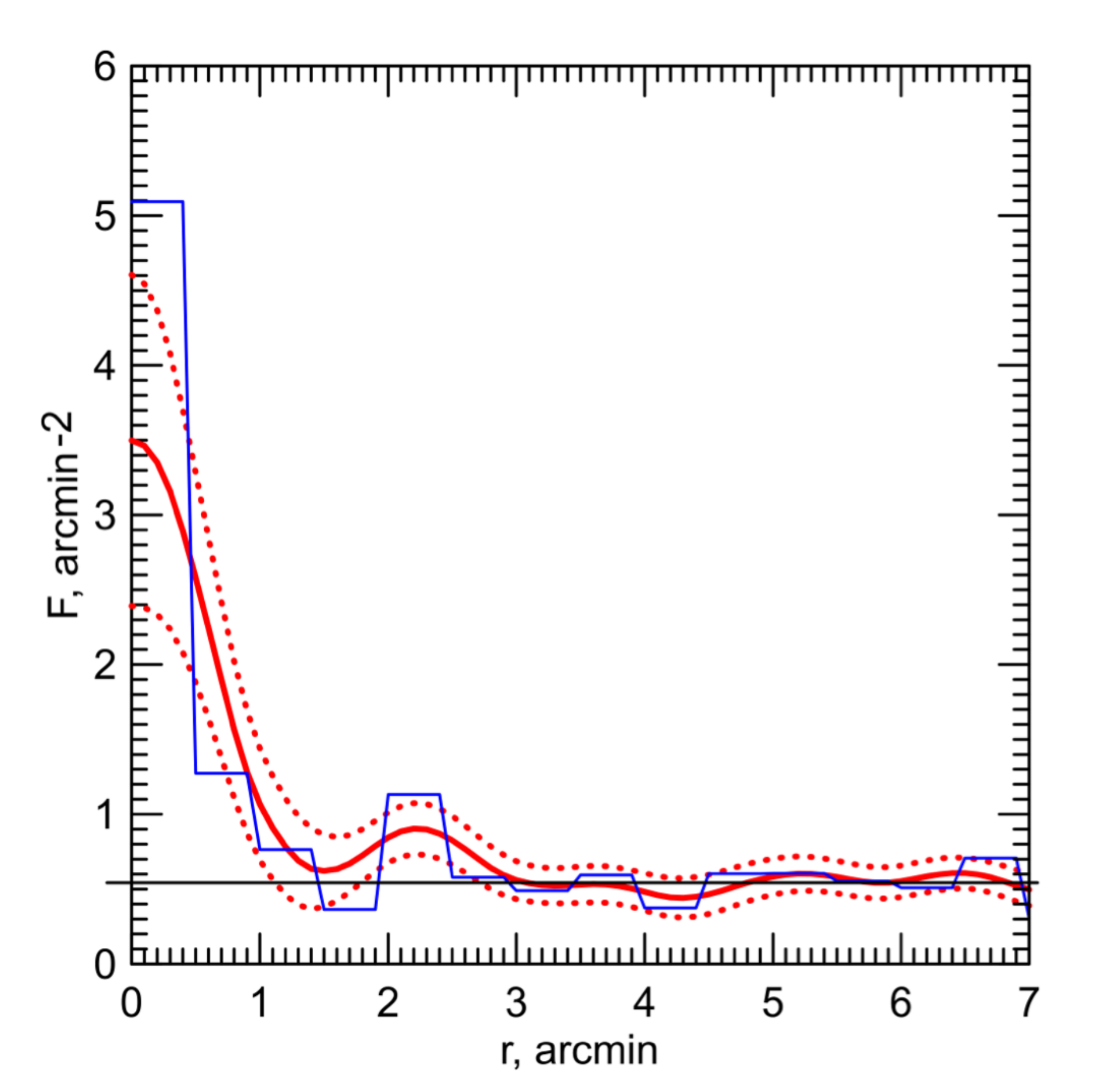}
  \includegraphics[scale=0.25]{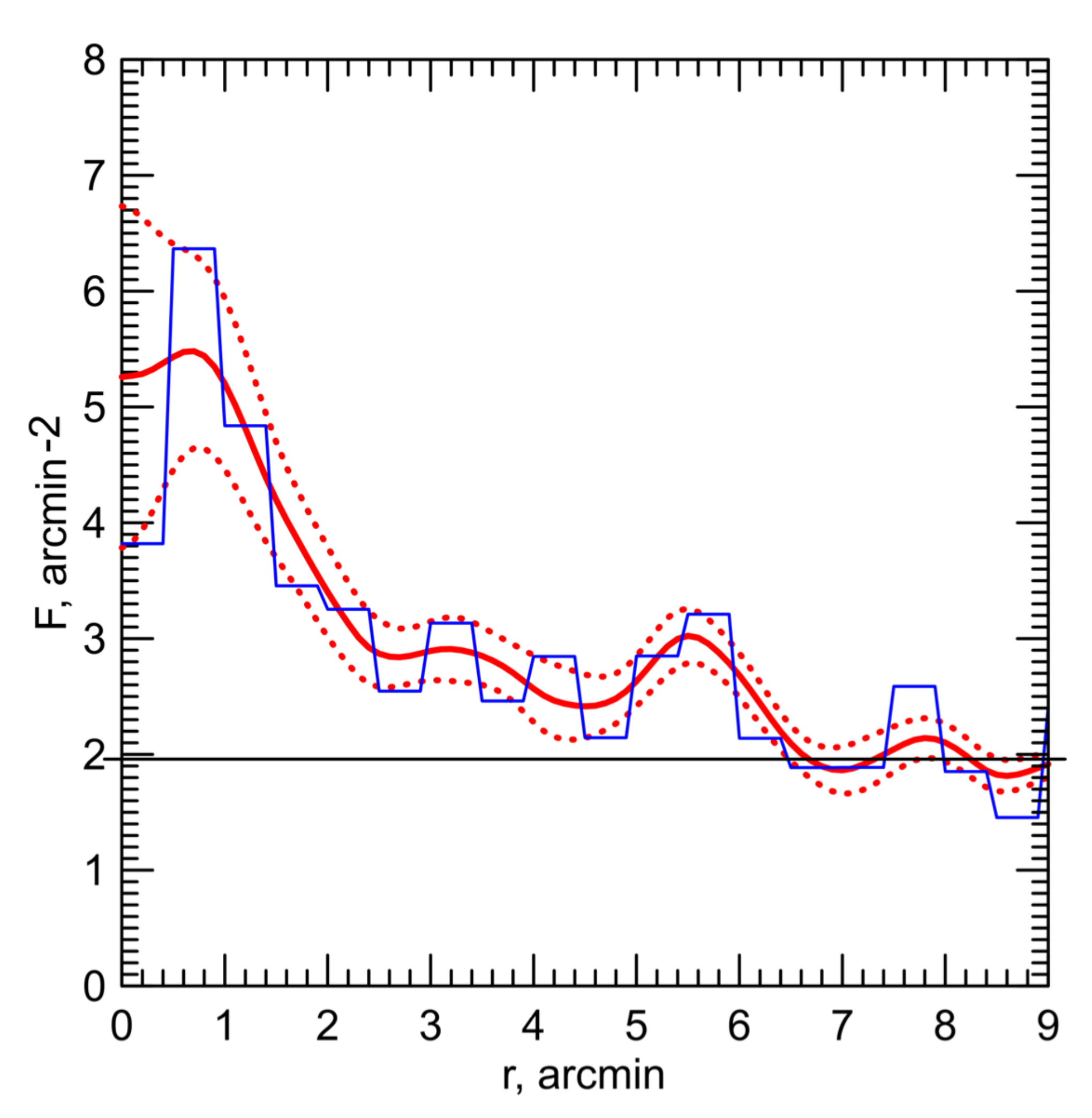}
  \includegraphics[scale=0.25]{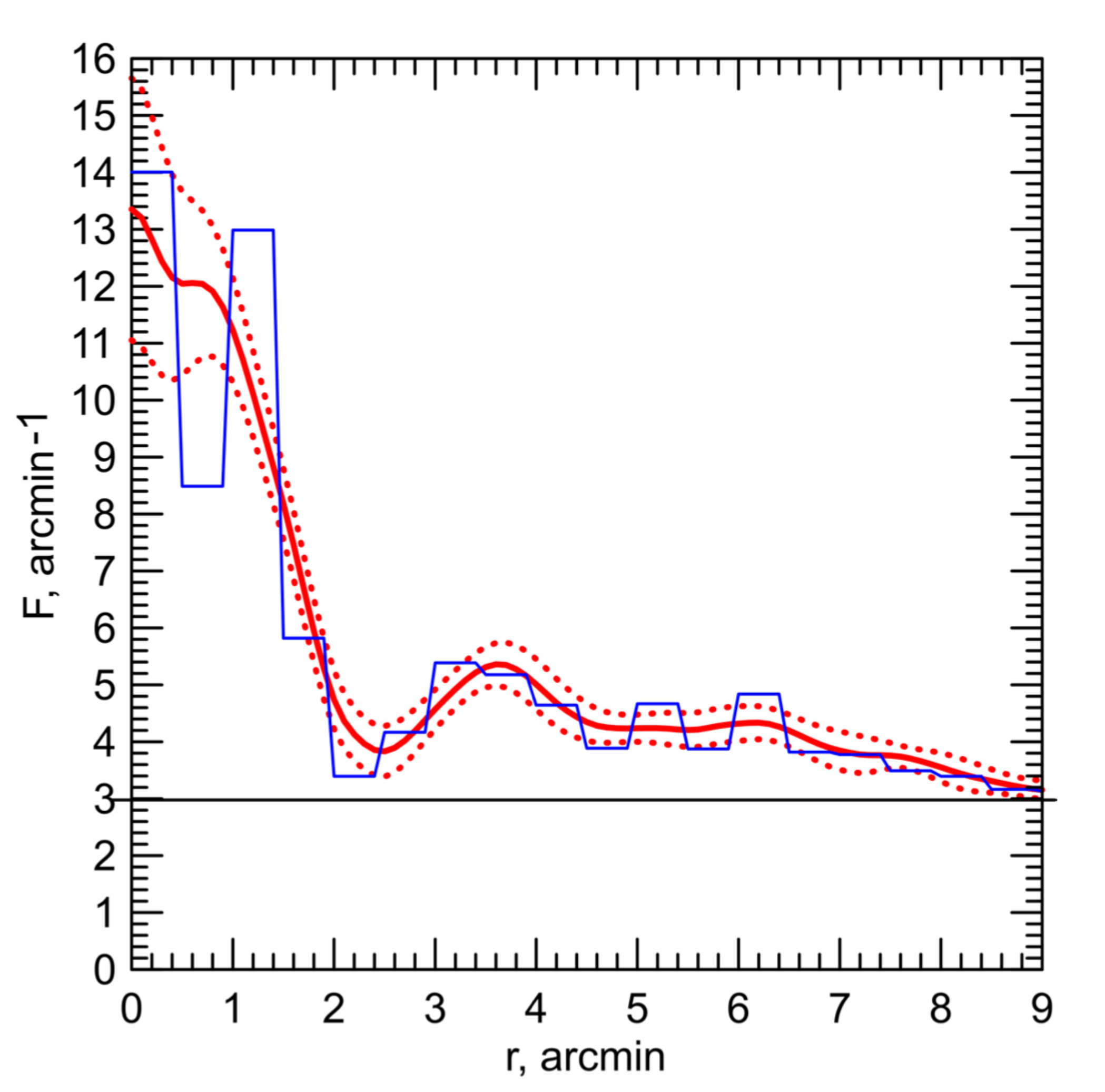}
  \includegraphics[scale=0.25]{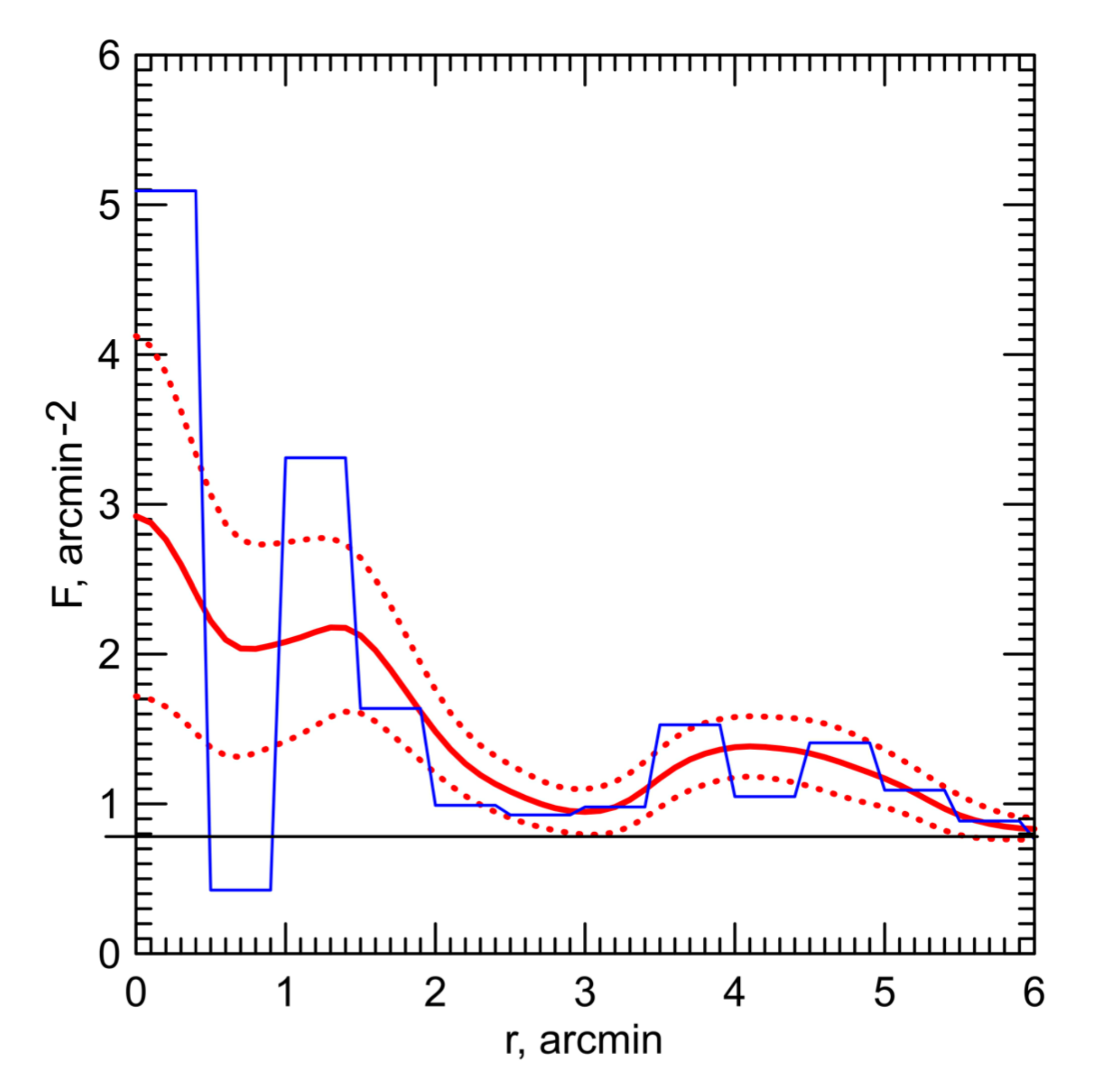}
  \caption{Surface radial profiles. 
  Bottom: Ruprecht 11 and ESO489SC01. Top, from the left to the right: NGC~2215, NGC~2354, and Haffner~22. The solid black line indicates the level of density background. The red curve is the cluster density profile, while the dashed red curves comprise the associate uncertainties. The blue line, finally, is the density histogram.}
   \end{figure*}
   
\noindent   
\begin{description}
\item $\bullet$ Inspection of the CMD of NGC~2354 in the upper right panel of Fig.~4 reveals three prominent features: a cluster main sequence with a turnoff (TO) point at $V\sim13.5$  and a handful of scattered red clump stars at $V\sim11.5$, a second, thick, well-populated main sequence (MS, colour-coded in red, with a TO at $V\sim19.5$ that looks like the MS for an intermediate-age/old stellar population, and a plume of blue stars, colour-coded in blue, in the magnitude range 16--18.5 that resembles a young stellar population. The last feature is very similar to the blue plumes found in the directions of other clusters (Moitinho et al. 2006; Carraro et al. 2005) in the third Galactic quadrant.
\item $\bullet$ Although those three features are readily visible in the NGC~2354 field, that is not the case for other lines of sight. In the CMD for NGC~2215 (upper left panel) there is no blue plume at all. The cluster MS itself is clearly visible with a TO at V $\sim$ 12. In addition, there is a thick, faint MS with a TO at $V=18.5$, which, however, is less prominent than in NGC~2354.
\item $\bullet$ The CMD for the field centred on Ruprecht~11 (middle right panel) shares the same features as NGC~2354, although displaced in magnitude. The thick, faint, old population MS has a TO at $V\sim20$ and a blue plume in the {\it V} range 17--19. Cluster members themselves, if any, are not easy to detect. There may be a red clump at $V\sim13$ and a TO at $V\sim15$. We return to that point later.
\item $\bullet$  The CMD of Haffner~22 (middle-left panel) presents a different story. Cluster stars are prominent with a MS TO at $V\sim16$ and a conspicuous red clump at $V\sim13.5$. The MS is visible despite severe field star contamination. There appears to be a plume of blue stars, but much brighter ($V\sim12-16$) and scattered than in previous cases. Some may be blue straggler stars (BSS), which would imply no blue plume in the field. However, typical BSS do not spread over such a large magnitude range (De Marchi et al. 2006). While one cannot exclude that some are BSS, they are not expected to outnumber young field stars. Finally, the thick, old MS does not have a clear TO, only a sort of sharp color cut-off at $(B-V)\sim0.6-0.7$.
\item $\bullet$ The CMD for stars in the ESO489~SC01 field (lower panel) is completely different, with none of the previous features present. There is no evidence of a MS, and there are only a few bright stars detached in magnitude from the bulk of the stars that can have produced the impression of a star cluster. We therefore conclude that ESO489SC01 is a chance alignment of a few bright stars, and not a physical star cluster.
\end{description}

\section{Cluster fundamental parameters}    
Described here is the methodology of deriving fundamental parameters for program clusters using star counts in tandem with photometric analysis. Star clusters are by definition over-densities of stars that produce distinctive features in a colour-magnitude diagram (CMD). But star over-densities do not necessarily imply the existence of physical clusters, while the opposite may also be true. Some star clusters do not produce evident over-densities relative to the density of stars in the general Galactic field (Carraro 2006).      
 
\subsection{Star counts and cluster spatial properties}
Star counts for each cluster field were made from the V-band images by generating surface density maps from our photometric catalogues. Maps of surface density were derived using
a kernel estimator (see Seleznev et al. (2010), Carraro \& Seleznev (2012), and the detailed description of the method given by Silverman (1986)), with the kernel half-width chosen to
diminish the effect of density fluctuations and to avoid, for example, other density peaks inside a cluster. Maps are shown in Fig.~5 using 3 arcmin as kernel half-width value, with different colours indicating different densities.

Density values are shown in the vertical color bar on the right side of each map, and the axes show distance from the field center in arcmin. North is up and east is to the left, as in Fig.~1. The maps were used to determine the cluster centres, which are listed in Table~4. The table contains the adopted limiting magnitude $V_{lim}$ for the map generation (column ~1), the derived center co-ordinates in degrees (columns 2 and 3), and the estimated radius in
arcmin (column 4). The new center co-ordinates can be compared with the values in Table~1.

The center of each cluster was determined visually from the peak density established by the 
locus of iso-density lines surrounding the maximum, and is indicated by yellow straight lines in Fig.~5. The limiting magnitude $V_{lim}$ was determined from the analysis using maps for different $V_{lim}$ in combination with inspection of the CMDs to determine the magnitude range where a cluster appears most clearly. The map for that selected value of $V_{lim}$ was chosen as the best representation for each cluster. A kernel halfwidth $h=3$ arcmin was used for all clusters except NGC 2354.
This is because $h=3$ arcmin  (upper mid panel in Fig.~5) clearly  misses  the cluster center for NGC 2354 due to its irregular star distribution, which is clearly visible  in the $h=1 arcmin$ map (upper right panel in Fig.~5) and 
in the density profile of the cluster (see below). This latter has been eventually used to estimate the cluster center.
It is worth noting that, with a kernel estimator, the field of the map is smaller than the field covered by the data by one smoothing length ($h$).

The coordinates for each cluster center were used to compute surface density profiles using the kernel estimator method (see detailed description in Seleznev (2015)). The method is more powerful than standard discrete binning, since it weighs the contribution of adjacent bins
using a kernel. The density profiles are shown in Fig.~6, where the vertical axis indicates surface density in units of arcmin$^{-2}$ and the horizontal axis is distance from the cluster center in arcmin. Recall that the limit in distance is determined by the kernel halfwidth (it is 1 arcmin for all clusters) and distance from cluster center to the nearest field border, since the cluster center rarely coincides with the center of the covered field.

Red solid lines show the computed density profiles, with red dotted lines depicting the 2$\sigma$ confidence intervals obtained from the smoothed bootstrap estimate method (Merritt \& Tremblay 1994; Seleznev 2015). The blue polygonal line in each case is the density histogram, while black horizontal lines represent visual estimate for the background densities. They were established as follows. If the density profile dropped to a nearly constant value, the background level was drawn so that the square of the areas above and below that level were about the same. The cluster radius was then estimated as the abscissa of the point of intersection of the density profile with the background density. The associated uncertainty corresponds to the separation between the intersections of confidence interval lines with background density.

\begin{table*}
\tabcolsep 0.2truecm
\caption{Revised coordinates of the cluster centers, cluster radii, and cluster fundamental parameters,}
\begin{tabular}{cccccccc}
\hline
Cluster & V$_{lim}$ & RA & Dec & radius & Age & Dist. & E(B-V)\\
\hline
 & [mag] & [hh:mm:ss.s] & [dd:mm:ss.s] & [arcmin] & [Gyr] & [kpc] & [mag]\\ 
\hline\hline
ESO489SC01 &  18& 06:05:11.8  &  -26:45:11.4  &   $\ge$6 (?)   & & &0.04$\pm$0.02   \\
NGC~2215     &  16 & 06:20:44.5  &  -07:17: 06.6 &   $\ge$7.6    & 0.8$\pm$0.15& 0.95$\pm$0.05 &0.16$\pm$0.02 \\
Ruprecht~11   &  18 & 07:07:21.5  &  -20:48:19.2  &   $\ge$9  &  1.5$\pm$0.5 & 187$\pm$0.13 & 0.17$\pm$0.02\\
NGC~2354     &  14 & 07:14:01.8  &  -25:39:29.9  &    3.1$\pm$0.4 (?) &  1.5$\pm$0.2 & 1.75$\pm$0.15 & 0.18$\pm$0.02\\
Haffner~22     &  16 & 08:12:27.9  &  -27:54:37.8  &    6.7$\pm$0.2 & 2.5$\pm$0.3 & 3.05$\pm$0.30 & 0.20$\pm$0.02\\
\hline\hline
\end{tabular}
\end{table*}

\noindent
We underline here that in the case of Haffner 22 and NGC 2354 the size of the fields may simply be too small, so that the estimated radius would refer only to the cluster core. As noted by Kholopov (1969), most star clusters consist of a concentrated nu- clear region surrounded by a lower density corona, typically larger than the fields imaged for each cluster. To establish the full extent of each cluster would therefore require estimating star densities be- yond the fields surveyed here (see, for example, Turner et al. 2008). It seems therefore more conservative to consider the values in Ta- ble 4 as lower estimates of radius.    

In the specific case of NGC 2354, the map in Fig.~5 (upper mid panel) reveals a second density maximum about 7 arcmin from the center reported in Table~4. If the density profile for NGC 2354 is derived by considering only to
$r=7$ arcmin (as in Fig.~6), the second maximum is not detectable. If it is considered as part of the cluster (which is highly probable), the cluster radius might be as large as $R\gtrsim 12$ arcmin.

  \begin{figure*}
  \includegraphics[scale=0.75]{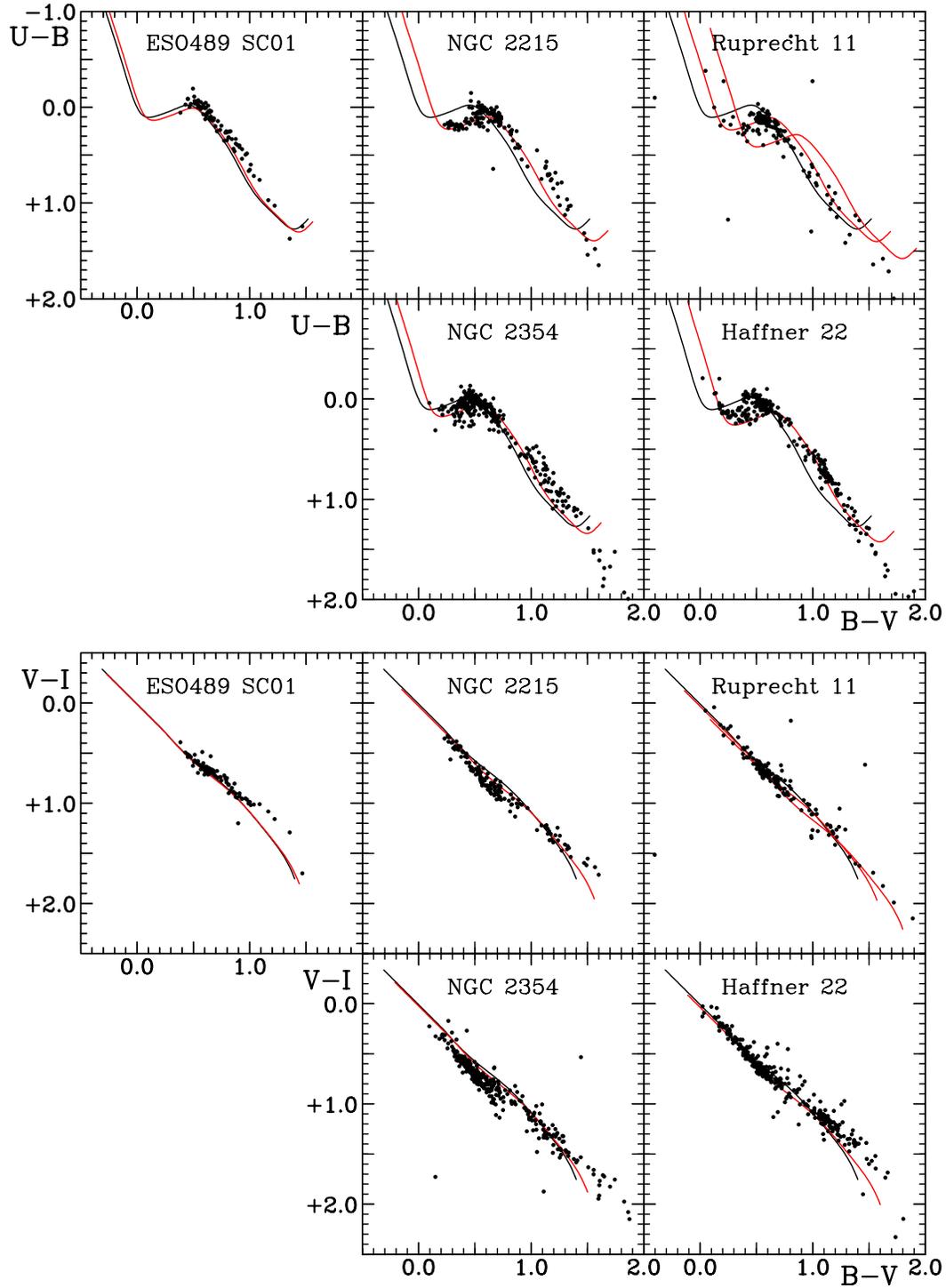}
  \caption{{\it U--B} vs. {\it B--V} and {\it V--I} vs. {\it B--V} color-color diagram for stars in our fields with UBVI photometry brighter than $V=16$. The intrinsic (black) and reddened (red) relations for solar composition main sequence stars are plotted, with reddenings as noted in the text.}
  \end{figure*}

\begin{figure*}
  \includegraphics[scale=0.35]{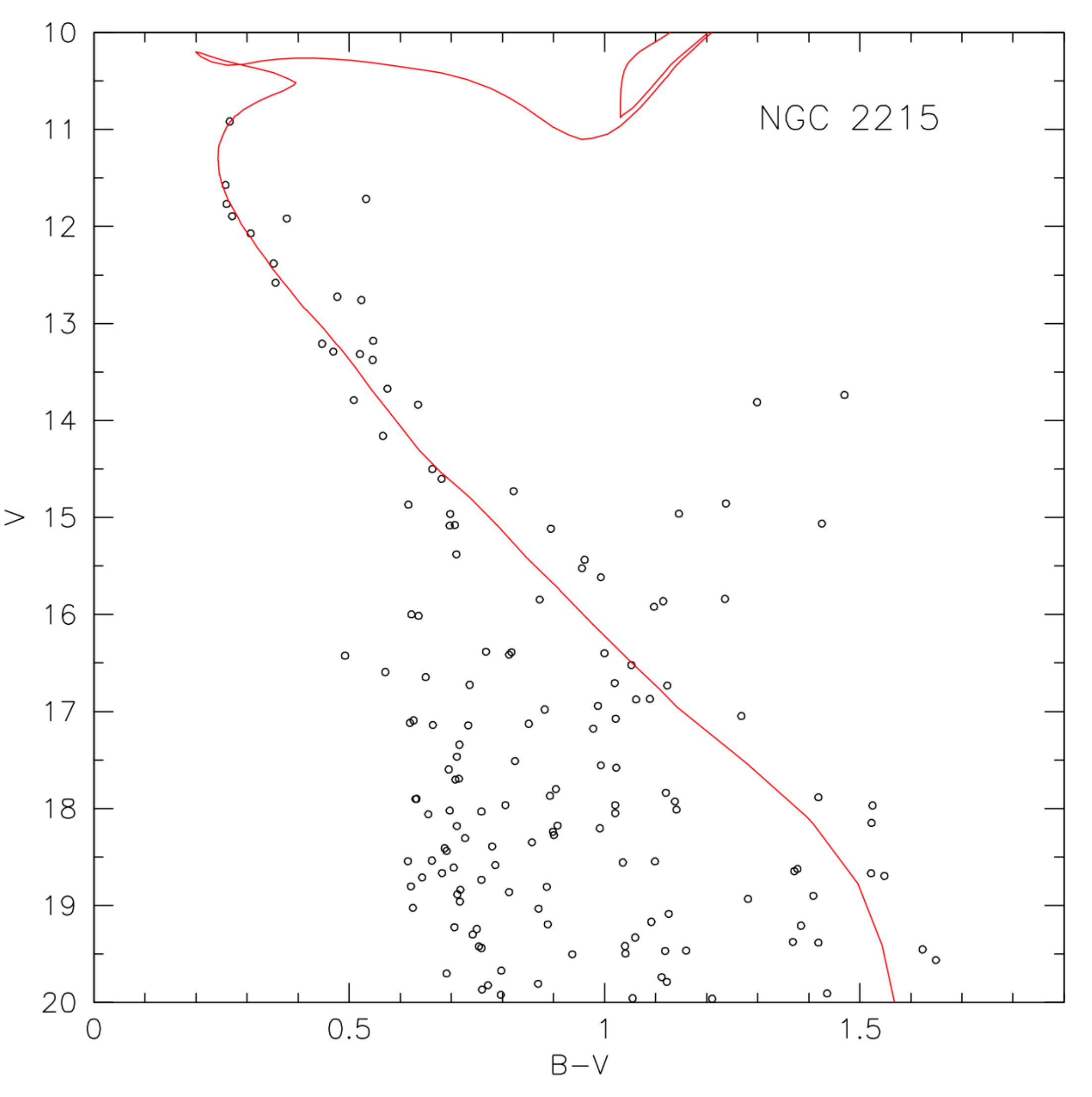}
   \includegraphics[scale=0.35]{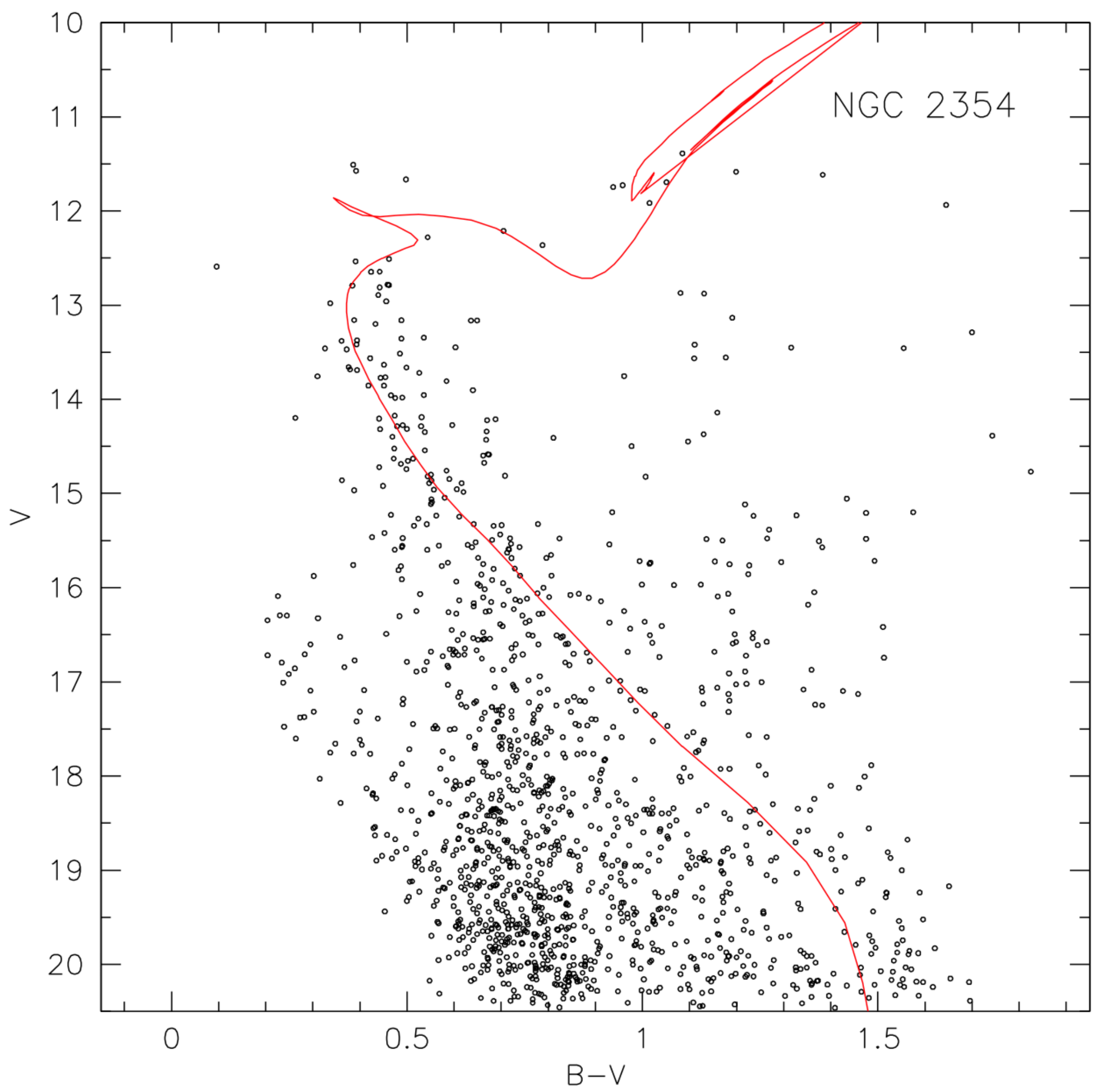}
  \includegraphics[scale=0.35]{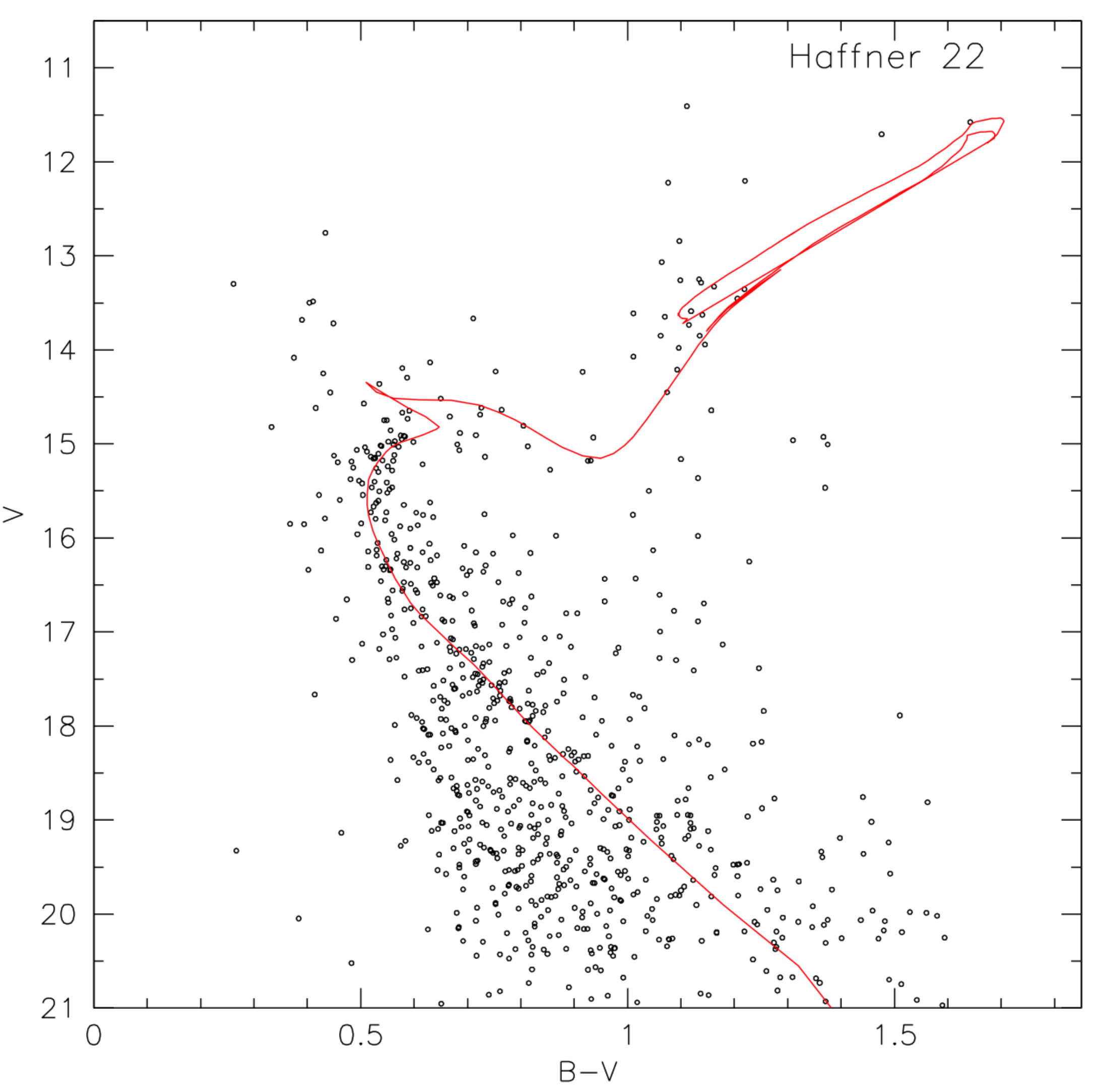}
  \includegraphics[scale=0.35]{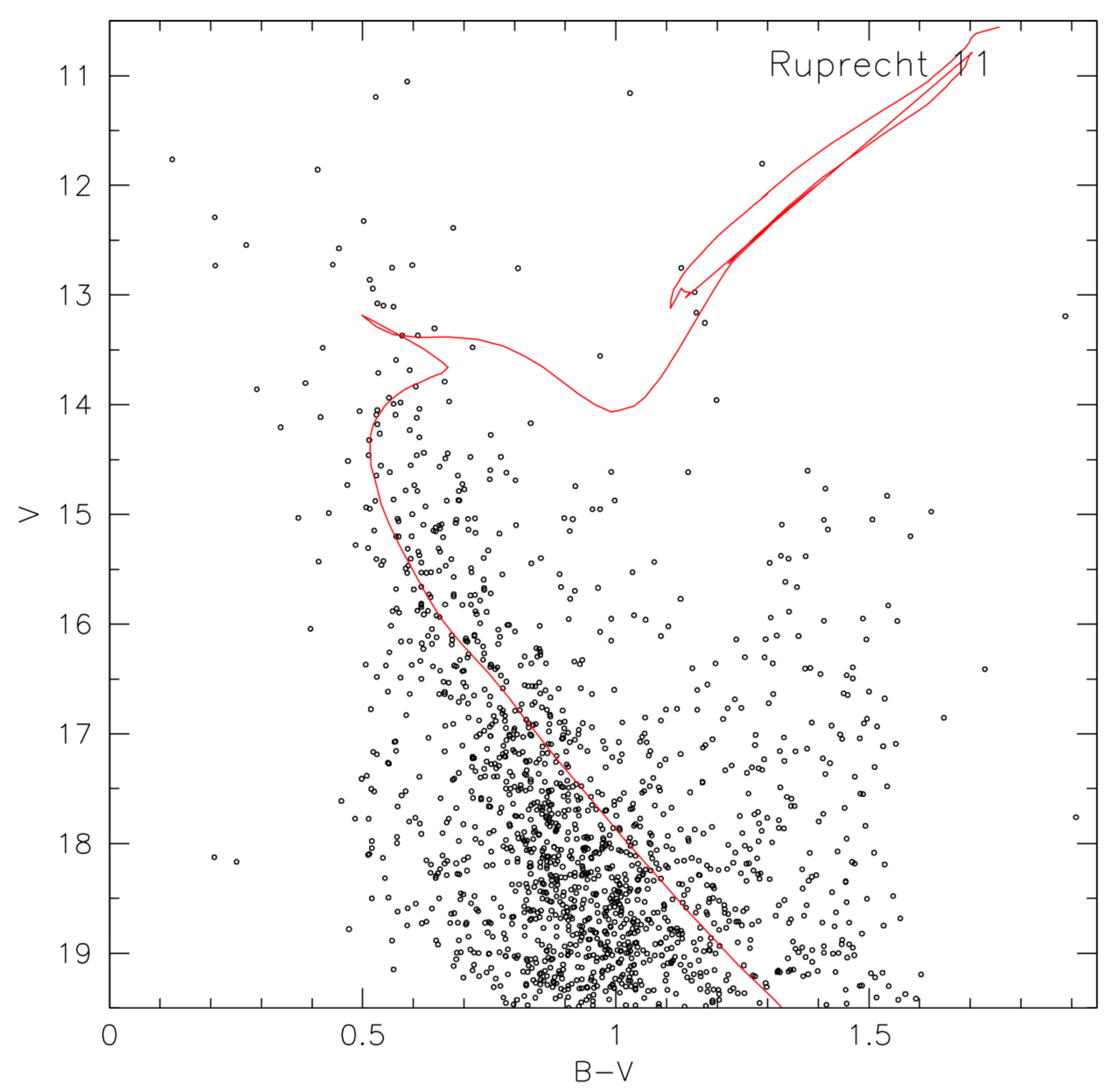}
  \caption{Isochrone solutions. Top panels: NGC~2215 and NGC~2254, from left to right. 
  Bottom panels: Haffner 22 and Ruprecht 11, from left to right. Fitting parameters are listed in Table~4.}
  \end{figure*}

\begin{figure*}
  \includegraphics[scale=0.6]{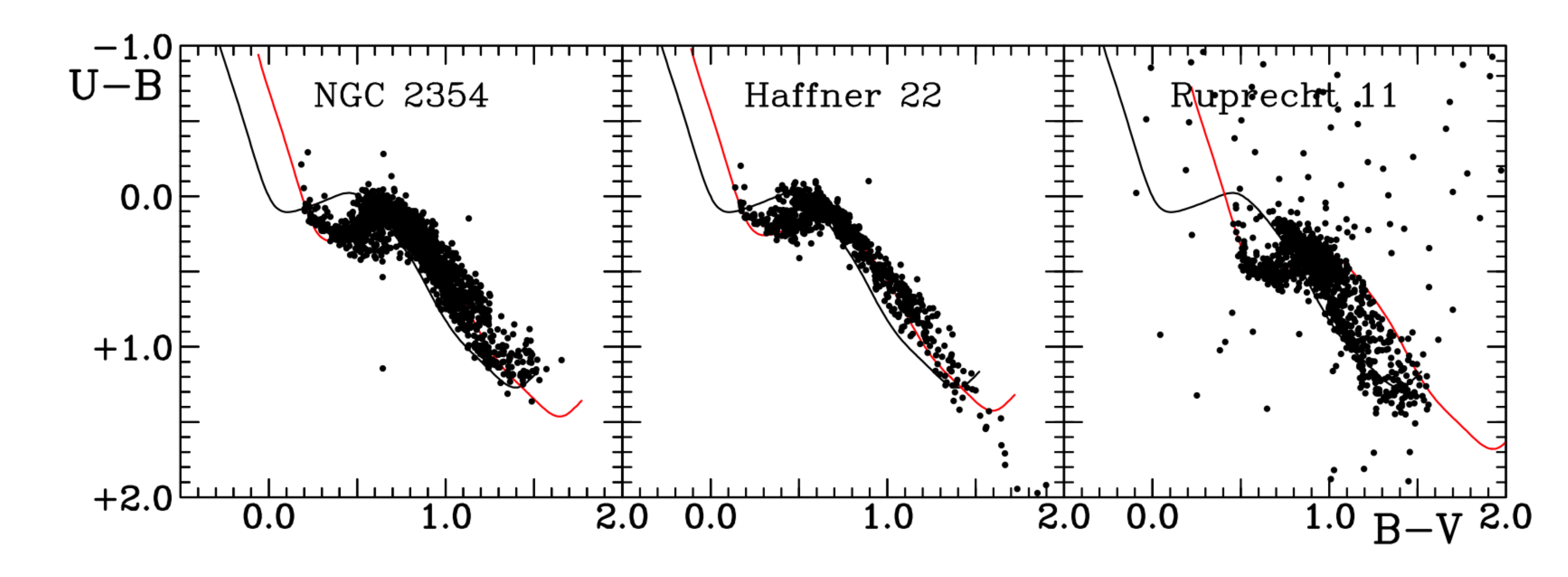}
  \caption{Similar {\it U--B} vs. {\it B--V} color-color diagrams as in Fig. 7 for stars in the magnitude ranges which comprise the blue plumes in NGC~2354, Haffner~22, and Ruprecht~11. Lines are as in Fig. 7, although the reddenings differ as noted in the text.}
  \end{figure*}
  
\subsection{The reddening law in the third Galactic quadrant}
The availability of photometry in four different bands ({\it UBVI}) and 2MASS observations for each cluster field make it possible to establish reddenings for each cluster using two-color data. The reddening law in several open clusters lying along segments of the Milky Way examined here was studied forty years ago by Turner (1976), and more recently by Turner (2012), who established localized reddening laws described by E$_{U-B}$/E$_{B-V} = 0.77$, which was adopted here. The corresponding derived values of $R_V$ range from 2.75 to 3.20, but a recent study of the reddening law (Turner et al. 2014) implies that a value of $R_V = 2.9$ would be more suitable. The reddening law in the Kron-Cousins system appears to be more stable, with a value of E$_{V-I}$/E$_{B-V} = 1.257$ being appropriate for all values of E$_{U-B}$/E$_{B-V}$ (Turner et al. 2011). For clusters lying away from the Galactic plane, a reddening law of slope E$_{U-B}$/E$_{B-V} = 0.83$ with $R_V = 2.8$ would seem to be more appropriate (Turner et al. 2014), but the reddening in such fields is typically so small that any differences do not affect the results significantly.

Color-color diagrams in {\it UBV} and {\it BVI} for each field are shown in Fig.~7, where the intrinsic relations in {\it UBV} are from Turner (1996) and those in {\it BVI} are from Turner et al. (2011). The data were limited to stars brighter than {\it V} = 15 in order to limit uncertainties in the colors to about $\pm0.01$. The {\it UBV} relation for GK-type stars incorporates the main sequence relation for stars of Pleiades (roughly solar) metallicity (Turner 1979) as well as the relation for ``zero-age zero-rotation'' AF-type stars (see Turner 2012), so is ideal for detecting, although not calibrating, the ultraviolet excesses expected for cooler stars lying away from the Galactic plane (Wildey et al. 1962). The intrinsic colors of giant stars (LC III) are also indistinguishable from those of dwarfs of late spectral type, so the relation is also useful for establishing the reddening of old open clusters.

It is important to point out that interstellar reddening and extinction does not characteristically increase linearly with distance, as is sometimes assumed. Rather, it increases in direct proportion to the amount of dust, or possibly hydrogen, along the line of sight (Turner et al. 2011, 2014). In such conditions all photometrically-measured stars in a field contribute valuable information on the reddening of both cluster and field stars.

The data of Fig.~7 provide insights into the reddening in each field. In the direction of ESO489~SC01, for example, only a few late-type stars are unreddened, and most exhibit ultraviolet (UV) excesses, typical of directions away from the plane. Most solar-metallicity stars exhibit a reddening of E$_{B-V}=0.04 \pm0.02$, with the reddening and extinction occurring in dust clouds close to the Galactic plane.

Most stars lying in the direction of NGC~2215, by contrast, are significantly reddened, on average by E$_{B-V}= 0.16 \pm  0.02$, with very few unreddened foreground stars in the population. Stars lying in Ruprecht~11 exhibit more complicated reddening. Many are unreddened foreground objects, but at some distance differential reddening takes over. Clumps of stars reddened by E$_{B-V}=0.17$ and E$_{B-V}=0.40$ are evident, with the main group being less reddened. A full differential reddening analysis may be needed to sort out cluster parameters. The stars lying towards NGC~2354 are similar in some respects, in displaying differential reddening, although there does appear to be a main group reddened by E$_{B-V}=0.18 \pm0.02$. Finally, stars lying towards Haffner~22 are less complicated, with a main group displaying reddenings of E$_{B-V}=0.20 \pm0.02$.
         
\subsection{Star cluster fundamental parameters}
With the results of the previous section on hand, it is possible to highlight star cluster
sequences in the CMDs, and then estimate their fundamental parameters by means of isochrone fitting. Use was made of the most recent suite of Padova models (Bressan et al. 2012), in conjunction with a radial selection of stars using the cluster center and radius estimated in Table~4. The results are shown in the panels of Fig.~8, with comments on a cluster-by-cluster basis. It is important to stress that the CMD sequences are sometimes difficult to fit because of field star contamination and differential reddening. To make the fitting solutions less confusing in Fig.~8, only the best-fit isochrones are shown. They were always superposed by eye. Associated uncertainties were also estimated by eye, adjusting each isochrone in color and magnitude until an acceptable fit was no longer possible.\\

\noindent
{\bf NGC~2215}\\
NGC~2315 is a difficult cluster to study, since it is sparsely populated, without an obvious RGB clump and with a poorly-defined TO region. Yet the cluster emerges clearly as an overdensity above the general field, and its MS is obvious (upper left panel in Fig.~8). Metallicity is difficult to infer without spectroscopy, but the colors of the few FG-type members fit the reddened Pleiades relation well in Fig.~7, so are unlikely to be far from solar. Adoption of the Fitzgerald et al. (2015) estimate ([Fe/H] $\simeq$ --0.3) leads to a very similar parameter set: E$_{B-V}=0.20 \pm0.05$, (m-M) = 10.2$\pm$0.2, and an age of 800$\pm$150 Myr. Solar metallicity would be more appropriate, however, for a cluster less than 1 kpc from the Sun (Lepine et al. 2011, Fig.~3).\\

\noindent
{\bf NGC~2354}\\
The present study yields a similar parameter set to a previous study by Claria et al. (1999).  
The cluster is 1.75$\pm$0.15 kpc distant with an age of 1.5$\pm$0.2 Gyr and E$_{B-V}$ = 0.18$\pm$0.02 (see top right panel in Fig.~8). With a prominent red giant clump and a TO about 1.4 magnitudes fainter, the cluster cannot be as young as the age of about 250 Myr suggested by Kharchenko et al. (2013). If the Carraro \& Chiosi (1994) calibration is used, NGC~2354 would be as old as IC~4651 (1.5 Gyr). Consequently, the $\sim$4 kpc distance inferred by Kharchenko et al. (2005) can also be ruled out.\\

\noindent
{\bf Haffner~22}\\
Haffner~22 (lower-left panel in Fig.~8) is a beautiful, intermediate-age, star cluster that closely resembles the more famous clusters NGC~2420 and NGC~2506. The curved TO is extremely clear, and the RG clump as well, although quite sparse. The present study yields the parameter estimates: age of about 2.5$\pm$0.3 Gyr, reddening of E$_{B-V}$ = 0.20$\pm$0.02, and distance of $\sim3.05\pm0.30$ kpc. Haffner~22 is therefore one of the few clusters located above the formal Galactic plane ($l=0^{o}$) in the third quadrant, where the disk is warped significantly towards negative declinations.\\

\noindent
{\bf Ruprecht~11}\\
According to star counts Ruprecht~11 shows the strongest density contrast with the surrounding field among our sample of clusters. Yet the CMD is very difficult to interpret, likely because of the evidence for differential reddening in the direction of the cluster (see above). Nevertheless, it does appear to be a cluster, provided the group of red stars at $V \sim 13-13.5$ is the RGB clump (see lower-right panel in Fig.~8). Main-sequence and isochrone fitting then yields the following parameters for Ruprecht~11: age of 1.5$\pm$0.5 Gyr, reddening of E$_{B-V}$ = 0.17$\pm$0.02, and distance of 1.87$\pm$0.13 kpc.\\

\noindent
{\bf ESO489~SC01}\\
No fit was made for the ESO489~SC01 star distribution since there is no evidence for a cluster lying in this direction. The CMD and color-color plots leave the impression of a lightly-reddened star field lying toward the Galactic halo.\\

\noindent
The results of our analysis are summarised in Table~4.

\section{Analysis of the stellar fields}
It is now possible to analyse the surrounding star fields along the lines of sight under investigation, by interpreting the various features highlighted in Section~4.

\subsection{The blue plumes}
Blue plumes are present in the background of three clusters, as noted above: NGC~2354, Ruprecht~11, and Haffner 22. No such feature is detected in the field of ESO489~SC01, while a weak background group of yellow dwarfs is visible in the field of NGC~2215. It is necessary to establish if the stars in each blue plume are young or old, or if they are dispersed in distance or spatially confined.

For that we consider in each CMD the magnitude strip that encompasses blue plume stars: $16\leq V \leq19$, $12\leq V \leq 17$, and $17\leq V \leq 19$ for NGC~2354, Haffner~22, and Ruprecht~11, respectively. The stars are plotted in {\it UBV} color-color diagrams in Fig.~9 to establish their reddening distribution and approximate spectral types, from which can be inferred an approximate age. Dwarf stars occupy the bluest region of the diagram ({\it B--V} bluer than 0.5 for NGC~2354 and Haffner~22, and bluer than 0.75 for Ruprecht~11). The solid black line is the intrinsic relation described earlier, while the red line is the same relation reddened to fit the bluest plume stars. The location of the blue branch does not depend on metallicity (Carraro et al. 2008a). The color excesses of the bluest stars correspond to reddenings E$_{B-V}=$ 0.25, 0.20, and 0.53 for NGC~2354, Haffner~22, and Ruprecht~11, respectively.

\begin{figure*}
  \includegraphics[scale=0.6]{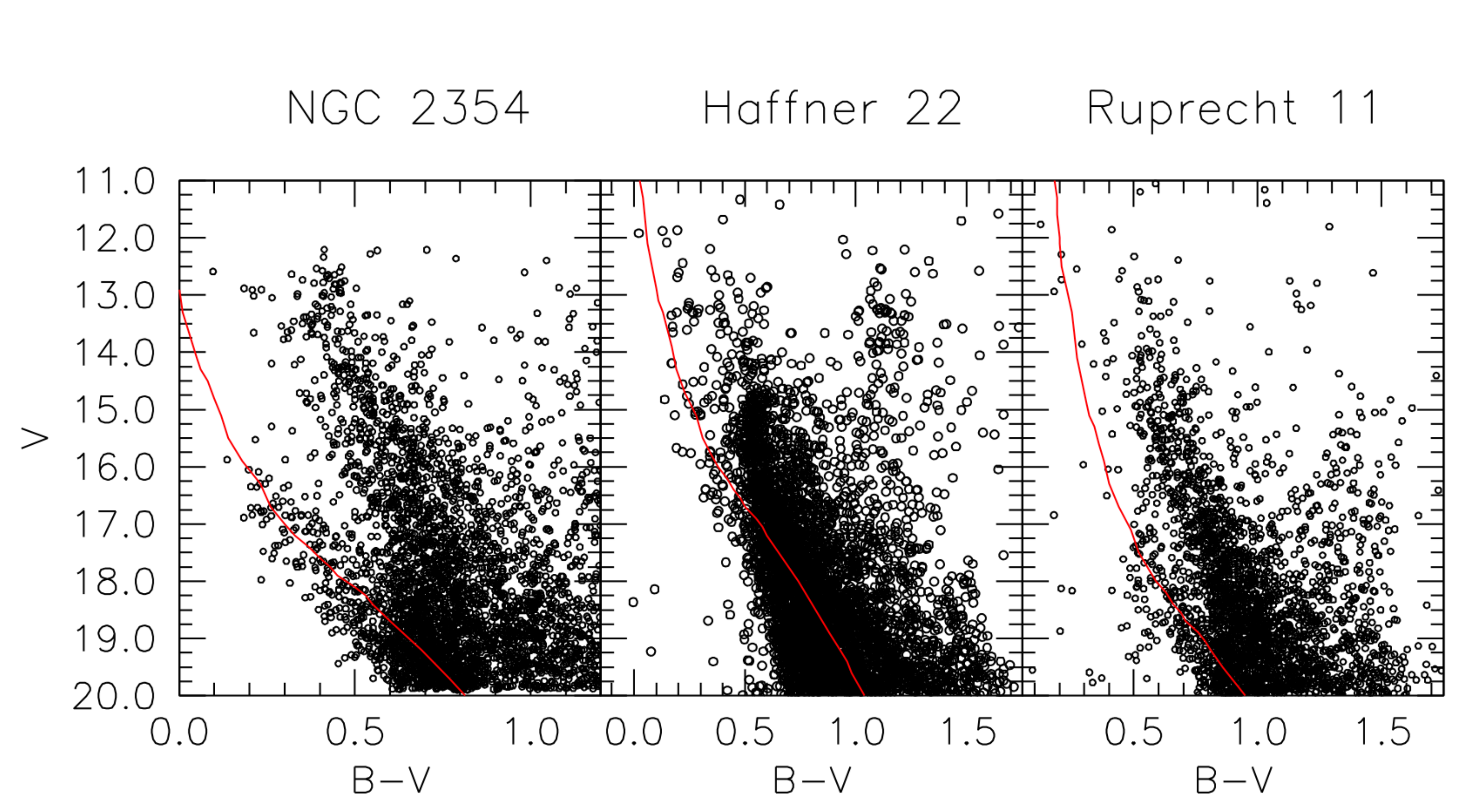}
  \caption{Distance solution for the blue plumes in NGC~2354, Haffner~22, and Ruprecht~11}
  \end{figure*}

From the data of Fig.~9 we infer that the earliest spectral types are B8 for NGC~2354, B9 for  Haffner~22, and B4 for Ruprecht~11, implying ages for blue plume stars in the direction of Ruprecht~11 as young as 50 million years, and slightly older, up to 70-80 million years, for those in the lines of sight to NGC~2354 and Haffner 22. Such ages are typical of stars in young open clusters that are preferentially located close to the spiral arm where they formed.

As noted earlier, the blue plumes in NGC~2354 and Ruprecht~11 appear to correspond to a spatially confined population, while blue plume stars in Haffner~22 are much more scattered, which would mean they are distributed over a larger range of distances. The mean distances from the Sun are estimated by superimposing a ZAMS on the stars in the CMDs, as illustrated in Fig.~10. That yields (m-M) = 15.5$\pm$0.2 for NGC~2354, (m-M) = 11.2 to 14.30$\pm$0.2 for Haffner~22, and (m-M) = 16.4$\pm$0.1 for Ruprecht~11. Therefore, in the lines of sight towards NGC~2354 and Ruprecht~11 young stars appear to peak in their distribution at distances of 9.0$\pm$0.8 kpc and 8.2$\pm$0.4 kpc, respectively, whereas blue plume stars lying towards Haffner~22 are dispersed over distance of 1.2--5.5$\pm$0.5 kpc. Such extreme Galacto-centric distances challenge modern Galactic models that adopt a sharp disk cut-off at 12--14 kpc (Carraro et al. 2010), and confirm beyond any reasonable doubt that the young Galactic disk is significantly warped at these Galactic longitudes.

\subsection{The faint, intermediate-age main sequence}
Attention can now be focused on the last prominent feature identified in the CMDs of NGC~2354, NGC~2215, and Ruprecht~11, a thick, faint MS typical of an intermediate-age, possibly metal poor, stellar population. The feature was not detected in ESO489~SC01 or in Haffner~22, although in the latter case the TO may simply be hidden because it is confused with the cluster itself and with the MS of foreground stars, located between the Sun and the cluster.

The direction where the population is best visible is that towards NGC~2354. Its MS is broad  in color for various reasons, among which are variable reddening across the field, different distances for the stars, and contamination by other Galactic components (thick disk and halo) sampled by the line of sight to the cluster. The same is true for NGC~2215 and Ruprecht~11. In the case of NGC~2215, Fitzgerald et al. (2015) suggest that this population is about 5 Gyr old for a metallicity as low as [Fe/H] = --0.8, which would then represent material from the Galactic thick disk, with little to do with the Perseus arm, where a much younger population is expected.

Sequences such as these are routinely found in the Galactic anticenter, and are normally considered to be part of the Monoceros ring, a structure that might represent the leftover of an in-plane accretion event (Conn et al. 2012 ). Such a narrow MS is not expected to be produced by the general Galactic disk population when sampling either along the plane, where one can find stars at any distance, or off-plane, where one does not expect to find many thin disk stars.

An alternative view has been proposed by V\'azquez et al. (2008, 2010) and Carraro et al. (2010, 2015), who suggest that this population is made of old thin disk stars that follow the Galactic warp. As in the case of the young thin disk, the warp introduces the effect that the line of sight intersects and eventually crosses the disk completely, producing the impression of a population confined in space (see Carraro 2014, Fig~9 for an illustration of the effect).

It should be stressed that stars belonging to this population are visible in the color-color diagrams in Fig.~9. They produce a {\it helmet}-like (or {\it hump}-like) structure where stars of FG spectral type are expected. That happens at 0.5 $\leq V \leq 0.7$ for NGC~2354,
at 0.4 $\leq V \leq$0.6 for Haffner~22, and at 0.5 $7\leq V \leq$ 0.9 for Ruprecht~11. It is an indication that such stars have ultraviolet excesses, namely they are bluer in {\it U--B} than the intrinsic relation for solar-metallicity stars, and therefore metal-poor. It confirms earlier findings by our group (Carraro et al. 2008a). Yet it is difficult to estimate what the metallicity is precisely, since we are looking at a population spread in space at different distances, which might include old thin disk, contamination from thick disk, and even halo stars. More importantly, the derivation of photometric metallicity using the ultraviolet excess index $\delta(U-B)_{(B-V)_o=0.6}$ requires precise photometry and homogenous (in terms of stellar population) samples (see Carraro et al. 2008b). At best it can be noted that the component is more metal poor than the Sun.

There is no obvious way to derive an age estimate for this component. In Carraro et al. (2008a) we superimposed two extreme isochrones to the same population (but located at $l =244^{o}$, $b=-8.0^{o}$) and provided only an indication of a possible age range, concluding that what we are seeing at low latitude is probably thick disk population. Here, we try a more empirical approach. In Fig.~10 we compare the CMD of stars in the NGC~2354 field with that for the intermediate-age star cluster NGC~2158 (Carraro et al. 2002). The latter is a well-known,
rich, intermediate-age ($\sim$ 2 Gyr) cluster that is metal-poor ([Fe/H = --0.3), located in the second Galactic quadrant. It is also known to suffer from significant variable reddening that makes it an ideal comparison with the stellar population studied here, which we tentatively assume to be composed mostly of old thin-disk stars, spread by reddening, and more metal-poor than the Sun.
 
For the purpose of the comparison, NGC~2518 stars were displaced in color and magnitude by $\Delta(B-V) =-0.4$ and $\Delta V =1.5$, respectively, to position the cluster sequence where the old, faint population in NGC~2354 is located. Although contamination is strong, it can be argued that, if one eliminates blue plume stars from the NGC~2354 CMD, the MS of NGC~2158 would then reproduce and match qualitatively well the faint MS of NGC~2354. It is not a conclusive argument, but only an attempt to provide a simple and direct comparison with similar stellar populations located in other regions of the disk, where the warp is less important. Given the similarity with the MS of NGC~2158, the faint MS would mostly be old, thin-disk population. 

\section{Discussion and conclusions}
Analysed here are the CMDs of five star fields located in the third Galactic quadrant and centered on the catalogued open clusters NGC~2215, NGC~2354, Haffner~22, Ruprecht~11, and ESO49~SC01. Star counts and photometry are used to discuss their nature and properties, and to provide estimates of fundamental parameters. Those for NGC~2215 and NGC~2354 agree with results of previous studies, while those for Haffner~22 and Ruprecht~11 are new. We cannot provide solid confirmation for the reality of the latter, but the former is an intermediate-age open cluster with a conspicuous RGB clump. Finally, we argue that ESO489~SC01 is not a star cluster, but only a chance overdensity projected toward the Galactic halo.

The main goal of the study was to analyse the additional features present in the CMDs, namely the existence of young populations in the field of NGC~2354, Haffner~22, and Ruprecht~11. In the case of Haffner ~22, this population is not spatially confined, but is distributed along the line of sight to the cluster, from 1.2 kpc from the Sun to as distant as 5.5 kpc, implying that the line of sight to Haffner~22 (at $b$=+3$^{o}$.37) crosses the local arm (Orion) only, and then misses the outer thin disk which starts to bend below the plane because of the warp at about 3--4 kpc from the Sun (Carraro et al. 2015, Vazquez et al. 2008).

The blue plume populations in NGC~2453 and Ruprecht~11, on the other hand, share the same properties. They describe a Galactic feature confined in space, located at about 8--9 kpc from the Sun, and about 1 kpc below the Galactic plane. That feature, according to our previous investigations, coincides with the location of the outer spiral arm, also known as the Norma-Cygnus arm (Vazquez et al. 2008). Disk warping renders it easier to detect the feature since the line of sight intersects the arm transversally, thus producing the effect of detecting a spatially confined structure (see Carraro 2014, Fig.~9, for a clear illustration of the effect).

Interestingly, no blue plume is seen in NGC~2215 at $l \sim 215, b \sim -10^{o}$, which would indicate that the young disk is not present at such low latitudes. That is in agreement with the warp structure as derived by clump stars (Momany et al. 2006), which predicts the maximum of the warp at larger Galactic longitudes ($l \sim 230-250$).

An additional comment can be made. From the analysis of color-color diagrams we find that the blue plume population in Ruprecht~11 ($b=-5.981^{o}$) is younger than those in NGC~2354 ($b=-6.792^{o}$) and Haffner~22 ($b=+3.377^{o}$). That may imply the existent of a vertical age gradient, with the direction towards Ruprecht~11  sampling a more centrally concentrated (and therefore younger) region of the warped Galactic thin disk.

Also noted is the presence of an older, metal-poor, stellar component in the fields of NGC~2354, NGC~2215, and Ruprecht~21. They would be old thin disk populations, similar to that in the open cluster NGC~2158 or Tombaugh~2, and therefore not older than 2-3 Gyr. It is, however, difficult to estimate the metallicity from photometry only. The population is present along the lines of sight in different proportions. It is quite strong towards NGC~2354 and Ruprecht~11, which are located relatively close in the third quadrant. Ruprecht~11 shows
a significantly larger reddening, and therefore the corresponding faint MS appears much fainter than in the case of NGC~2354. On the other hand, NGC~2215 is located much more distant from the warped Galactic plane than NGC~2354, which implies that the line of sight to the cluster intersects a lower and less dense portion of the disk. The same population may also be present towards Haffner~22. But in that direction we may be seeing only the northern edge of the disk. In such a situation we would not be able to see a distinct TO, since the line of sight, close to the Sun, is almost parallel to the formal Galactic plane, and stars are spread over a large distance range from the Sun to the cluster, and beyond, up to 5 kpc, as in the case of the blue plume stars. Beyond 5 kpc from the Sun, the line of sight would no longer intersect the disk, which is warped, and located at much lower latitudes.

Finally, this population is totally absent at the latitude and longitude of ESO489~SC01. In our opinion, future lines of research should focus on the derivation of stellar parameters of stars in both young and intermediate-age populations using spectroscopic data, to better assess their metallicity properties and nature.

\begin{figure}
  \includegraphics[width=\columnwidth]{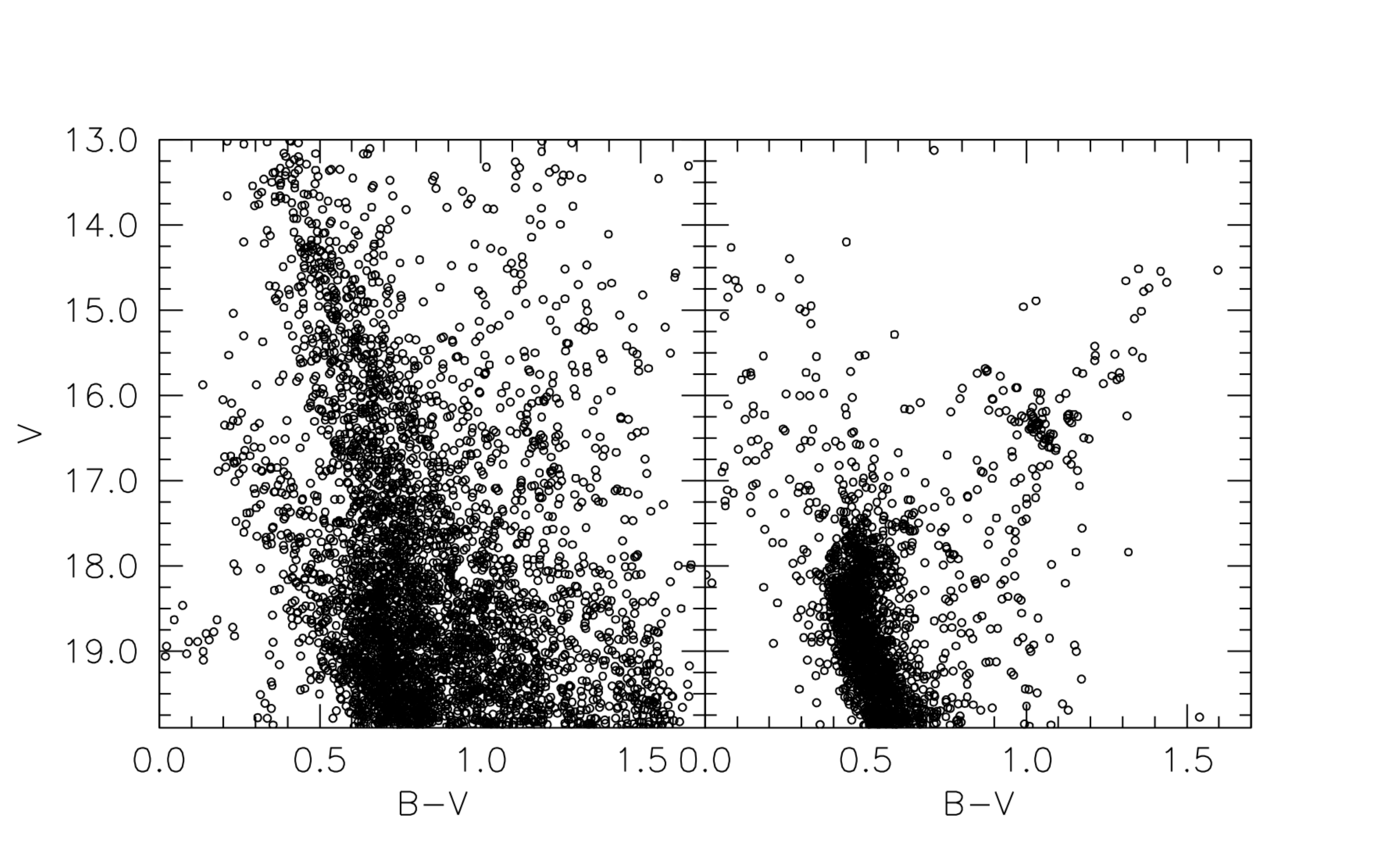}
  \caption{Comparison of NGC~2158 MS (right panel) with the intermediate-age, faint MS in the region of the star cluster 
  NGC~2354 (left panel).}
  \end{figure}
 
\section*{Acknowledgments}
The work of A.F.Seleznev was supported by the Ministry of Ed- ucation and Science of the Russian Federation (state contract No. 3.1781.2014.K, registration number 01201465056). The travel of G.Carraro was supported by Act 211 Government of the Rus- sian Federation, contract No. 02.A03.21.0006. G. Carraro acknowledges financial support from ESO DGDF program and from the UrFU Competitiveness Enhancement Program, that allowed him to visit Ural Federal University in Ekaterinburg, where the work was completed. G. Baume acknowledges financial support from CONICET PIPs 112-201101-00301 and from the ESO visitor program that allowed a visit to ESO premises in Chile, where part of the work was done. The authors are much obliged for the use of the NASA Astrophysics Data System, the SIMBAD data base and ALADIN tools (center de Donn\'{e}es Stellaires, Strasbourg, France), and of the WEBDA open cluster data base. This publication also made use of data from: (a) the Two Micron All Sky Survey, which is a joint project of the University of Massachusetts and the Infrared Processing and Analysis Center/California Institute of Technology, funded by the National Aeronautics and Space Administration and the National Science Foundation; (b) the AAVSO Photometric All-Sky Survey (APASS), funded by the Robert Martin Ayers Sciences Fund.



\end{document}